\documentclass[aps,pra,reprint,groupedaddress]{revtex4-1}

\usepackage{graphicx}
\usepackage{dcolumn}
\usepackage{bm}
\usepackage{amsmath,amssymb}
\usepackage{physics}
\usepackage{dsfont}
\usepackage{color,soul}
\usepackage[utf8]{inputenc}
\usepackage{qcircuit}
\usepackage{svg}

\usepackage{rotating}

\usepackage[english]{babel}

\makeatletter
\adddialect\l@en\l@english
\makeatother

\def\eye{\mathds{1}}

\newcommand{\add}[1]{#1}

\begin{document}


\title{Machine Learning Phase Transitions with a Quantum Processor}
\author{A.\,V.~Uvarov}
\email{alexey.uvarov@skoltech.ru}
\author{A.\,S.~Kardashin}
\email{andrey.kardashin@skoltech.ru}
\author{J.\,D.~Biamonte}
\email{j.biamonte@skoltech.ru}
\affiliation{Deep Quantum Laboratory, Skolkovo Institute of Science and
Technology, 3 Nobel Street, Moscow 143026, Russia}



\begin{abstract}
Machine learning has emerged as a promising approach to unveil properties of many-body systems. Recently proposed as a tool to classify phases of matter, the approach relies on classical simulation methods---such as Monte Carlo---which are known to experience an exponential slowdown when simulating certain quantum systems. To overcome this slowdown while still leveraging machine learning, we propose a variational quantum algorithm which merges quantum simulation and quantum machine learning to classify phases of matter.  Our classifier is directly fed labeled states recovered by the variational quantum eigensolver algorithm, thereby avoiding the data reading slowdown experienced in many applications of quantum enhanced machine learning.  We propose families of variational ansatz states that are inspired directly by tensor networks.  This allows us to use tools from tensor network theory to explain properties of the phase diagrams the presented method recovers.  Finally, we propose a nearest-neighbour (checkerboard) quantum neural network. 
This majority vote quantum classifier is successfully trained to recognize phases of matter with 99\% accuracy for the transverse field Ising model and 94\% accuracy for the XXZ model. These findings suggest that our merger between quantum simulation and quantum enhanced machine learning offers a fertile ground to develop computational insights into quantum systems.
\end{abstract}

\maketitle

\paragraph*{Introduction.}  The best contemporary algorithms to emulate quantum systems using classical computers suffer from an exponential slowdown in limiting cases. A recent approach is to apply machine learning, which offers new techniques for large-scale data analysis. In particular, machine learning was recently proposed as a tool to recognize phases of matter \cite{carrasquilla_machine_2017,van_nieuwenburg_learning_2017}. These methods still rely on Monte Carlo sampling (or alternative classical simulation methods) which suffers from the an exponential slowdown induced by the so-called sign problem. Independently, quantum algorithms have also been proposed as a platform for machine learning \cite{biamonte_quantum_2017,huggins_towards_2019,schuld_circuit-centric_2018,schuld_quantum_2019,havlicek_supervised_2019,biamonte_quantum_2017,schuld_introduction_2015,duan_quantum_2017,duan_a_optimal_2019,sheng_distributed_2017}. In addition, unlike classical algorithms, quantum simulators are predicted to simulate quantum systems efficiently \cite{georgescu_quantum_2014,bernien_probing_2017,barreiro_open-system_2011}. Here we merge quantum machine learning with quantum simulation, leveraging quantum mechanics to overcome two classical bottlenecks. Namely, we leverage quantum algorithms as a tool to simulate quantum systems by preparing states which are labeled and fed into a quantum classifier.  The later removes the data reading slowdown experienced in many applications of quantum enhanced machine learning.  While the former utilizes quantum simulators to avoid classical methods with known limitations.  We also replace the standard unitary coupled cluster ansatz found in implementations of the variational quantum eigensolver \cite{peruzzo_variational_2014} with families of tensor network ansatz states.

The idea of machine learning is to recognize patterns in data. Using machine learning techniques, one can analyze the phase diagrams of strongly interacting quantum systems and thus directly address system properties. In this approach, Monte Carlo samples of such systems are used as input data and classified using supervised \cite{carrasquilla_machine_2017,van_nieuwenburg_learning_2017} or unsupervised \cite{wang_discovering_2016} learning. This way, spin Hamiltonians of up to a few hundreds of entities can be studied. Nonetheless, for fermionic systems, the use of Monte Carlo methods is drastically restricted by the sign problem, leading to an exponential slowdown.

In the variational quantum circuits approach \cite{mcclean_theory_2016, wecker_progress_2015}, the quantum computer is required to prepare a sufficiently rich variety of probe states (or circuits, e.\,g.\, if the problem is to approximately compile a certain quantum gate \cite{khatri_quantum-assisted_2018}). This approach emerged in response to challenges to adopt quantum algorithms for existing hardware. A particular example, the variational quantum eigensolver (VQE), represents an implementation of variational quantum circuits which uses a quantum processor to prepare a family of states characterized by a polynomial number of parameters and minimizes the expectation value of a given Hamiltonian within this family. This approach is widely taken on small- and middle-size quantum computers \cite{peruzzo_variational_2014,omalley_scalable_2016,colless_computation_2018}. 

In this Letter, we propose a way around the sign problem using a quantum computer. To classify the phases of a given quantum Hamiltonian, we first prepare its approximate ground states variationally, and then feed them as an input to a quantum classifier. In this respect, there is no need in sampling of microscopic configurations with Monte Carlo based methods. Instead, the classifier has direct access to the quantum states \cite{havlicek_supervised_2019}, yielding thus an effective realization of quantum-enhanced machine learning. To make this algorithm realizable on near-term quantum computers, we propose preparing the quantum states using shallow tensor network based circuits. Numerical tests show that this technique allows the quantum classifier to correctly recognize phase transition in transverse field spin models.

\newpage





\paragraph*{Tensor network ansatz states for VQE.} VQE is a hybrid iterative quantum-classical algorithm used to approximate the ground state of a given Hamiltonian \cite{peruzzo_variational_2014}. It relies on preparing an ansatz state $| \psi(\boldsymbol{\theta}) \rangle$ by applying a sequence of quantum gates $U(\boldsymbol{\theta})$ and sampling the expectation value of a given Hamiltonian relative to this state, followed by a classical optimizer to minimize the energy, $\langle  \psi(\boldsymbol{\theta}) | H |  \psi(\boldsymbol{\theta}) \rangle$. Within the VQE method, we approach the ground state of a given Hamiltonian using tensor networks ansatz states.

We proceed with representing the Hamiltonian as a sum of tensor products of Pauli operators:
\begin{equation}
    \label{eq:pauli_basis1}
        H = \sum \mathcal{J}_{\alpha_1 \alpha_2 \dots \alpha_n} \sigma_{\alpha_1} \otimes \sigma_{\alpha_2} \otimes \dots \otimes \sigma_{\alpha_n},
\end{equation}
where $\alpha_i \in \{0, 1, 2, 3\}$ enumerate the Pauli matrices $\{ \eye, X, Y, Z\}$. With the decomposition shown in Eq. \ref{eq:pauli_basis1}, individual terms of $\langle  \psi(\boldsymbol{\theta})| H | \psi(\boldsymbol{\theta}) \rangle$ can be estimated and variationally minimized elementwise using a classical-to-quantum process. In each iteration one prepares the state $\ket{\psi(\boldsymbol{\theta})}$ and measures each qubit in the local $X, \ Y$, or $Z$ basis, estimates the energy and updates $\boldsymbol{\theta}$. This method can become scalable only if the number of terms in the Hamiltonian is polynomially bounded in the number of spins and the coefficients $\mathcal{J}_{\alpha_1 \alpha_2 \dots \alpha_n}$ are defined up to poly($n$) digits.

It is therefore not surprising that the performance of VQE crucially depends on the choice of the ansatz state. A common approach is to use the unitary version of the coupled cluster method, the unitary coupled cluster (UCC) ansatz \cite{shen_quantum_2017,omalley_scalable_2016,hempel_quantum_2018,dumitrescu_cloud_2018}. For interacting spin problems, the (non)unitary coupled cluster ansatz can be composed out of spin-flip operators \cite{mcclean_theory_2016,gotze_heisenberg_2011}. There is no known classical algorithm to efficiently implement this method, even when the series is truncated to low-order terms \cite{taube_new_2006}. In principle, a quantum computer could efficiently prepare this state, truncated up to some $k$-th order using the Suzuki--Trotter decomposition \cite{lanyon_towards_2010}. However, for a system of $n$ qubits it requires $\mathcal{O}(n^k)$ unitary gates, making this technique out of reach for the available quantum computers. Still, even if UCC is truncated to single and double interactions (UCCSD), it requires $9n^2$ operations and necessitates applying some optimization strategy that would remedy this problem \cite{cao_quantum_2018}.

\begin{figure}
    \centering
    \includegraphics[width=\linewidth]{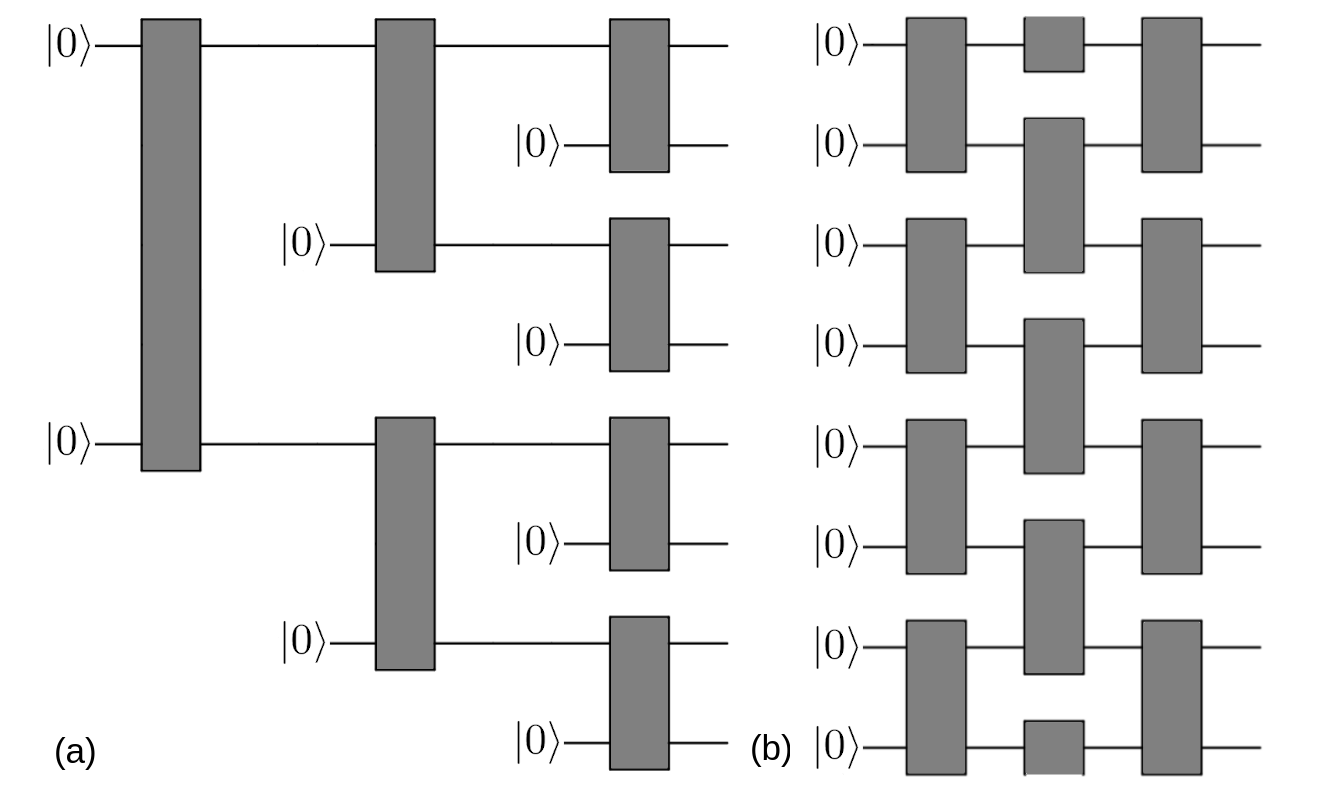}
    \caption{(a) Tree tensor network state. (b) Checkerboard tensor network state. In both cases, the quantum register is instantiated in the $\ket{0}^{\otimes n}$ state and subject to entangling gates. Black boxes indicate two-qubit gates specified by Eq. \ref{eq:two_qubit_block1} and Fig. \ref{fig:entangler}. Each block is parametrized independently. \add{Here and in all subsequent figures, quantum circuits are to be read left to right}.}
    \label{fig:tree}
\end{figure}

Instead, we test a number of shorter ansatz states inspired by tensor network states, namely (i) a rank-one circuit; (ii) a tree tensor network circuit (Fig.~\ref{fig:tree}a); and (iii) a family of checkerboard-shaped circuits (Fig.~\ref{fig:tree}b) with varying depth.


\begin{figure}
    \centering
    \includegraphics[width=0.8\linewidth]{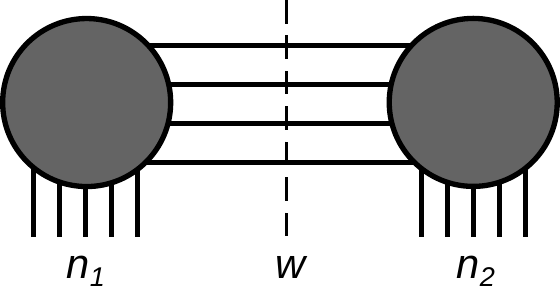}
    \caption{A quantum circuit can be treated as a tensor network state with all bonds having dimension 2. If a bipartition cuts $w$ wires, the total bond dimension is at most $2^w$, while the cut separates at most $w$ ebits of entanglement.}
    \label{fig:cut}
\end{figure}

In particular, these states differ in the amount of entanglement they can support. In general, quantum states of $n$ qubits can be represented by $n$-index tensors, while quantum circuits are embodied by tensor networks. Each quantum gate is seen as a vertex, and each string is an index running through $\{ 0, 1 \}$, while the maximum amount of entanglement is determined by the number of strings one needs to cut to separate the subsystems (see Fig.\,\ref{fig:cut}). Each string corresponds to at most one ebit of entanglement. An $n$-qubit state can contain at most $\lfloor n/2 \rfloor$ ebits of entanglement. To formalize the latter, suppose there exists a certain bipartition in the system that brings $n_1$ qubits to the first subsystem and $n_2$ qubits to the second. It is then possible to regroup the tensor state into a $2^{n_1} \times 2^{n_2}$ matrix. The rank of this matrix provides an upper bound to the amount of entanglement across this bipartition: a rank-$k$ state can support at most $\log_2k$ ebits of entanglement, i.e.\,when it is in the maximally entangled state.

{\it State preparation.} We first approximate the ground state of a Hamiltonian by rank-one states. One can prepare any unentangled state using $2n$ gates by subsequently applying $R_y$ and $R_z$ rotations to each qubit. This ansatz essentially matches the first order truncation of UCC.

Fig.~\ref{fig:tree}a illustrates a circuit implementing a {\it tree tensor network state}. 
\add{This state is prepared using the two-qubit parametric blocks subsequently described in detail. The preparation procedure is easier to explain when the number of qubits is a power of two. A block first entangles qubits number 1 and $(n/2 + 1)$. 
Then, in the next layer, two blocks act on qubit pairs $(1, n/4 +1)$ and $(n/2 + 1, n/2 + n/4 + 1)$.
Then, for each half of the register, we act in the same way, but now the blocks act on qubit pairs $(1, n/4 +1)$ and $(n/2 + 1, n/2 + n/4 + 1)$. Each half of the register is again divided into two halves. The pattern continues. For systems where the number of qubits is not a power of two, one can do this procedure up until the number of qubits involved is the closest power of two, and then make the last layer of operators incomplete.}
Such a structure enables long-range correlations but limits the entanglement entropy that a state bipartition can potentially have: any contiguous region can be isolated by $\mathcal{O}(1)$ cuts. It is easy to contract $\bra{\psi_{tree}} A \ket{\psi_{tree}}$ classically with $A$ being a local observable and $\ket{\psi_{tree}}$ being a tree tensor network state. 

\begin{figure}
    \centering
    \mbox{
    \Qcircuit @C=1.0em @R=1.0em {
           & \gate{e^{-i \tilde{\theta}_1 X}} & \multigate{1}{e^{-i \tilde{\theta}_3 {Z} \otimes {Z}}} & \gate{e^{-i \tilde{\theta}_4 Z}} & \qw \\
           & \gate{e^{-i \tilde{\theta}_2 X}} & \ghost{e^{-i \tilde{\theta}_3 Z \otimes Z}} & \gate{e^{-i \tilde{\theta}_5 Z}} & \qw \\
       }
    }
    \caption{Two-qubit entangler gate used in preparation of the states.}
    \label{fig:entangler}
\end{figure}
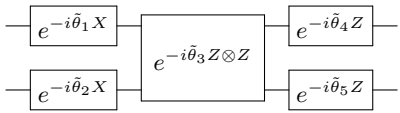

There is some freedom in choosing the two-qubit blocks that comprise the tree TN. In principle, one can implement any unitary in $SU(4)$ by using three controlled-NOTs and 15 single-qubit rotations \cite{vatan_optimal_2004}. However, throughout this work we used  two-qubit gates with fewer parameters $\bm{\tilde{\theta}}$ to simplify the optimization (Fig. \ref{fig:entangler}):
\begin{multline}
    \label{eq:two_qubit_block1}
    U(\boldsymbol{\tilde{\theta}}) = (R_z(\tilde{\theta}_5) \otimes R_z(\tilde{\theta}_4)) \ \circ \\ 
    \circ R_{zz}(\tilde{\theta}_3) \ \circ  \ (R_x(\tilde{\theta}_1) \otimes R_x(\tilde{\theta}_2)),
\end{multline}
where $R_z (\theta) = e^{i\theta Z/2}$, $\ R_x (\theta) = e^{i\theta X/2}$, and $R_{zz} (\theta) = e^{i\theta Z\otimes Z/2}$. Thus, a complete ansatz would have five free parameters per two-qubit block. \add{In total the tree tensor network ansatz features $n - 1$ blocks, yielding $5n - 5$ independent parameters.}

Remarkably, the block shown in Fig. \ref{eq:two_qubit_block1} is inspired by the parametrized Hamiltonian approach \cite{santagati_witnessing_2018} and the unitary operators used in the \add{quantum approximate optimization algotrithm} \add{(}QAOA\add{)} \cite{farhi_quantum_2014}. Of course, the ansatz with such blocks is weaker than it would be if each block implemented the entire $SU(4)$ group. However, such an ansatz would also have some redundancy as the ansatz gates are applied to a fixed $n$-qubit input state $\ket{0}^{\otimes n}$. 

In a {\it checkerboard tensor network}, the entangling blocks are positioned in a checkerboard pattern as shown in Fig.~\ref{fig:tree}b. In the following, we  impose periodic boundary conditions, meaning that the last qubit is linked to the first. For this ansatz, we also use the two-qubit entangling gate shown in Fig. \ref{fig:entangler}. \add{This way, the ansatz has $5 L \lfloor n/ 2 \rfloor$ independent parameters, where $L$ is the number of layers in the circuit.}

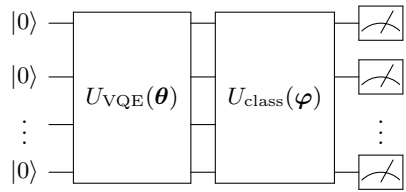
\begin{figure}
    \centering
    \mbox{
        \Qcircuit @C=1.0em @R=0.8em {
            & \ket{0} & & \multigate{3}{U_\mathrm{VQE} (\boldsymbol{\theta})}  & \multigate{3}{U_\mathrm{class} (\boldsymbol{\varphi})} & \meter  \\
            & \ket{0} & & \ghost{U_\mathrm{VQE} (\boldsymbol{\theta})}         & \ghost{U_\mathrm{class} (\boldsymbol{\varphi})}        & \meter \\
            & \vdots  & & \ghost{U_\mathrm{VQE} (\boldsymbol{\theta})}         & \ghost{U_\mathrm{class} (\boldsymbol{\varphi})}        & \vdots \\
            & \ket{0} & & \ghost{U_\mathrm{VQE} (\boldsymbol{\theta})}         & \ghost{U_\mathrm{class} (\boldsymbol{\varphi})}        & \meter \\
        }
    }
    \caption{Quantum circuit that implements the classifier. The first part prepares the VQE solution, the second one performs the classification. The assigned label is inferred from the measurements in the $Z$ basis. Both $U_{VQE}$ and $U_{class}$ have the checkerboard structure.}
    \label{fig:classifier_circuit}
\end{figure}

To isolate a region in a checkerboard tensor network state with $d$ layers, one has to cut at least $\mathcal{O}(d)$ bonds regardless of the region size (if the region is very small, one can also make a ``horizontal cut'' of $\mathcal{O}(L)$ bonds). Therefore, if we set the number of layers to be equal to $\lfloor \log_2 (n) \rfloor$, the upper bound on the entanglement scaling is equal to that in critical one-dimensional systems. To implement any maximally entangled state, that is, a state with the maximum possible amount of ebits, one needs to cut $\lfloor \frac{n}{2} \rfloor$ bonds.
\add{If the checkerboard ansatz is made with open boundary conditions, it needs at least $2 \lfloor \frac{n}{2} \rfloor$ gates to saturate entanglement. Periodic boundary conditions make the ansatz more powerful and lower this bound in half, to $ \lfloor \frac{n}{2} \rfloor$ layers.}

\paragraph*{Quantum classifier.} Not only can the checkerboard tensor network be used as a VQE ansatz but it also functions as a quantum majority vote classifier. 
\add{Each data point is a VQE solution: the parameters of the unitary gates are optimized in order to get the minimum energy $\bra{\psi(\boldsymbol{\theta})} H(h) \ket{\psi(\boldsymbol{\theta})}$, where $h$ is a parameter which determines the phase of the model.} 
Each VQE solution is labeled with ``0'' or ``1'' depending on whether the \add{model parameter} is above or below the phase transition point. 
We then prepare a circuit made of two parts (Fig. \ref{fig:classifier_circuit}). 
\add{The first part takes the blank qubit registry $\ket{000...0}$ and prepares the VQE solution in the form of an ansatz state. The second part takes this state as an input and applies a unitary $U_\mathrm{class} (\boldsymbol{\varphi})$.} 
We \add{then} measure the output of the circuit in the $Z$-basis. Let $q_0$ and $q_1$ be the total number of measurements in which more than half of the qubits are in ``0'' or ``1’’ states respectively. Finally, the classifier returns the predicted probability $p=q_1/(q_0+q_1)$ for the state belonging to the class ``1'' being equal to the probability that the majority of qubits vote ``1'', excluding ties. Fig. \ref{fig:ten-qubit_circ} in the Supplemental material shows the quantum circuit in more detail.

Let $\{ (\boldsymbol{\theta}_i, y_i) \}_{i=1}^{N_{train}}$ be the set of training data points and their labels, $y_i \in \{0, 1\}$. Let $p_i \in [0, 1]$ be the label predicted by the neural network. Then the logarithmic loss function is:
\begin{equation}
\label{eq:logloss}
    f = -\sum_{i=1}^{N_{train}} \left( y_i \log p_i + (1 - y_i) \log (1 - p_i) \right).
\end{equation}

To minimize $f$, we used the \add{simultaneous perturbation stochastic approximation (}SPSA\add{)} algorithm \cite{spall_multivariate_1992}. This algorithm estimates the gradient vector by computing a finite difference in random direction, then performs a gradient descent step. We optimized the log loss over 300 epochs, with both finite differences step size and learning rate starting very coarse and decreasing as $1/\sqrt{n_{epoch}}$, where $n_{epoch}$ is the epoch number.

\begin{figure}
    \centering
    \includegraphics[width=\linewidth]{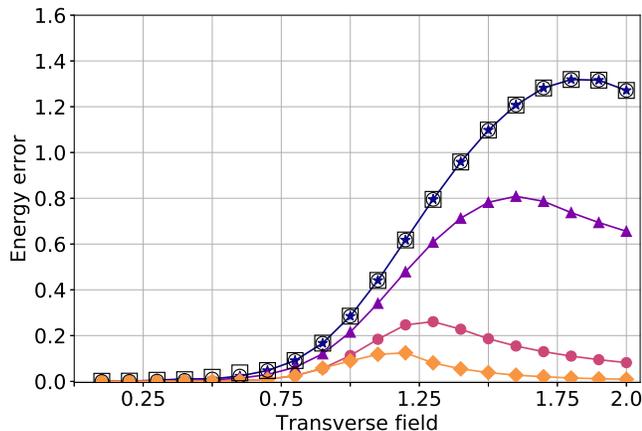}
    \caption{\add{Absolute value of the difference in energy between the exact solution and VQE solutions} for the transverse field Ising model. Hollow squares: rank-1 ansatz, hollow circles: tree tensor network, filled markers: checkerboard states ($\bigstar$: 1 layer, $\blacktriangle$: 2 layers, $\bullet$: 3 layers, $\blacklozenge$: 4 layers).}
    \label{fig:dE_Ising}
\end{figure}

\paragraph*{Numerical results.} To compare the performance of various VQE ansatz states, \add{we make numerical simulations of the quantum algorithm. In our simulation, the results of the measurements are given without noise and with perfect accuracy.} \add{W}e apply our quantum circuit of $n=10$ qubits to the transverse field Ising model (TFIM), which being exactly solvable \cite{lieb_two_1961,pfeuty_one-dimensional_1970} serves for testing purposes. This model is specified by the Hamiltonian
\begin{equation}\label{eq:ising}
    H_\mathrm{TFIM}=J\sum\limits_{i=1}^n\sigma_i^z\sigma_{i+1}^z+h\sum\limits_{i=1}^n\sigma_i^x, \; J>0, \; h>0,
\end{equation}
where $\sigma^\alpha_i$ is the Pauli matrix $\alpha$ acting on the $i$th spin, and we impose periodic boundary conditions $\sigma_{n+1}^\alpha=\sigma_1^\alpha$. 

\begin{figure*}
    \centering
    \includegraphics[width=\linewidth]{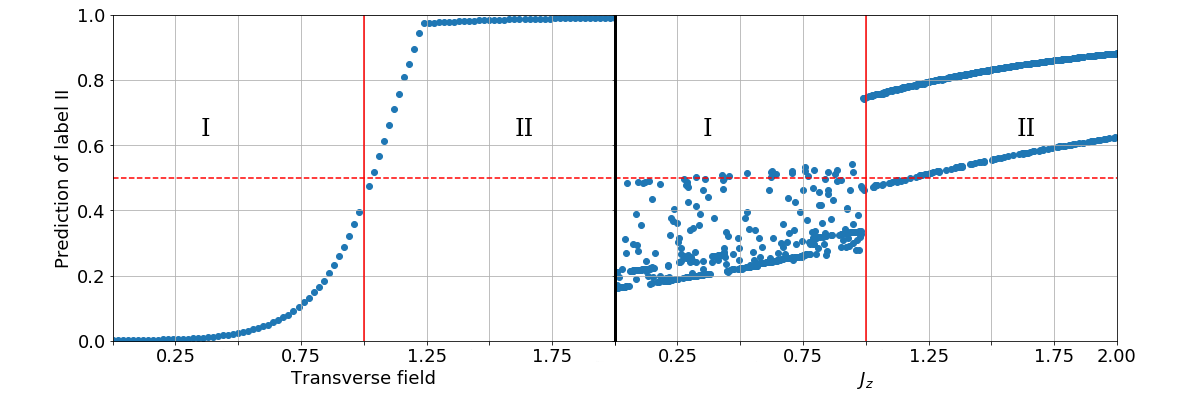}
    \caption{Left: predicted label of phase II as a function of magnetic field for transverse field Ising model. Right: predicted label of phase II as a function of $J_z$ for the XXZ model. Roman numbers denote the phases I and II of the models.}
    \label{fig:classification}
\end{figure*}

The ground state of TFIM is determined by the trade-off between Heisenberg exchange coupling, the first term in Eq. \ref{eq:ising}, favoring collinear orientation of magnetic moments in $z$ direction, and the Zeeman coupling to the transverse magnetic field, the second term. The latter has the tendency to flip the $z$-component, being thus the source of `quantum fluctuations' in the system. In a magnetically ordered state, a strong magnetic field $h\geq J$ destroys magnetic order even at zero temperature. This induces quantum fluctuations resulting in ground state restructuring, which is manifested by non-analiticity in the ground state energy of the quantum Hamiltonian (Eq. \ref{eq:ising}).

It is therefore intuitively clear that in the absence of magnetic field, $h=0$, or in the case of high spin polarization, $h=\infty$, the Hamiltonian is dominated by a single interaction, making the ground state disentangled so that even rank-one approximations could provide quantitatively correct results. Meanwhile, this is not the case at criticality, $h=J$, where the competition between the two mechanisms favors the formation of highly entangled ground state.


In our numerical implementation, we use Qiskit \cite{aleksandrowicz_qiskit:_2019} to simulate quantum circuits and the \add{limited Broyden–Fletcher–Goldfarb–Shanno method (}L-BFGS-B\add{)} to update the parameters during the classical step of the VQE, while all measurements are assumed ideal. We scan the values of $h$ from $0$ to $2J$. For $h=0$, the optimization process started from a random point, then each additional point started from the previous solution. To eliminate any obviously sub-optimal solutions, we also ran the scanning in the opposite direction and for each value of the field we keep the better result. \add{Without this ``double-sweeping'' procedure, spurious solutions appear in the vicinity of the phase transition: the solution for one phase remains a local optimum for some time after the parameter has moved to the other phase (see the Supplement).}

All ansatz states show an increase in error near the phase transition point  (Fig.\,\ref{fig:dE_Ising}). With the increasing depth, the checkerboard states show better approximation. It is interesting to point out that the low-depth ansatz states, namely rank-one, tree tensor network and checkerboard states of depth one, all show surprisingly similar results, which is associated with the relative simplicity of the Ising Hamiltonian. To feed the dataset into the classifier we prepare 100 data points using VQE with four-layered checkerboard ansatz states. This data was then shuffled and split into the training set (80\%) and the test set (20\%). Strikingly, the accuracy of the prediction achieves 99\%. The outcome of the quantum classifier is presented in Fig.~\ref{fig:classification} (left).


Another exactly solved model which we use to test our classifier is the antiferromagnetic $XXZ$ spin chain with the Hamiltonian:
\begin{equation}
    \label{eq:xxz_heis}
    H = \sum_{i=1}^n \left[J_\perp\left(\sigma^x_i \sigma^x_{i+1} + \sigma^y_i \sigma^y_{i+1}\right)
    + J_z \sigma^z_i \sigma^z_{i+1}\right]. 
\end{equation}
From a physical perspective, Eq. \ref{eq:xxz_heis} corresponds to a uniform exchange coupled system with a uniaxial anisotropy specified by $J_z$. At $|J_z| < J_\perp$, this model is in the XY, or planar, phase which is characterized by algebraic decay of equal-time spin-spin correlation functions. In the regime $J_z > J_\perp$ the Hamiltonian corresponds to the antiferromagnetic Ising state. The system undergoes a Berezinsky--Kosterlitz--Thouless type phase transition at $J_z = J_\perp$ is \cite{franchini_introduction_2017}. At the phase transition point, the ground state has the highest nearest-neighbour concurrence and a cusp in nearest-neighbour quantum discord \cite{dillenschneider_quantum_2008}.

This model has symmetry with respect to rotations in the $xy$ plane, as well as spin-flip symmetry. This fact allows us to augment the training data. Given a VQE approximation, we can create another, equally good approximation by applying a rotation or spin-flip. The structure of the VQE ansatz is conserved in the sense that the new states are produced by the same quantum circuit with different control parameters (see Supplemental Material for more details). In total, we produce 4000 data points. 

Despite being more subtle, the phase transition in the XXZ model is also correctly learned by the classifier, yielding correct labels on 94.6 \% of test data (Fig. \ref{fig:classification}, right). In this case, we have added two more layers to the classifier circuit to increase the accuracy. \add{Notably, the plot for the XXZ model looks less uniform than the plot for the Ising model. This is partially connected to the data augmentation procedure described in the Supplement. However, it is unclear why the behavior is abruptly changed at the phase transition point.}

\paragraph*{Conclusions \add{and discussion}.} We proposed a method of classifying phases of matter using a quantum machine learning algorithm paired with quantum simulation. In our numerical tests we achieved 99\% accuracy for the transverse field Ising model with a 4-layered classifier and 94\% accuracy for the XXZ model with a 6-layered classifier. The quantum classifier works intrinsically with quantum data, providing advantage over the classical methods based on Monte-Carlo sampling. \add{Manipulating quantum states explicitly on a classical computer would require an exponential amount of memory in the worst case. However, it would be interesting to compare our model with a classical model that would work with efficiently contractible states like matrix product states (MPS) or multiscale entanglement renormalization ansatz (MERA) states \cite{vidal_entanglement_2007}}.

\add{It is a nontrivial fact that the Ising model required fewer layers than the XXZ model. In the transverse field Ising model, the magnetization $\sum \langle \sigma_x^{(i)} \rangle$ as a function of magnetic field clearly points at the location of the phase transition points. This implies that the phases of the model are ``easy'' to classify. In the XXZ model, the transition at $J_z=1$ is a transition between a paramagnetic and an antiferromagnetic phase \cite{franchini_introduction_2017}. Neither of these phases shows spontaneous magnetic moment in absence of an external field, making it somewhat harder to discern the two phases. 
}

\add{The number of layers that can be used both for VQE and for the classifier is limited by several factors. Firstly, too long a circuit would be hard to implement in near-term quantum hardware. Second, the runtime of the classical optimization grows with the number of parameters in the cost function, which is $O(nL)$ for the checkerboard ansatz. Finally, McClean et al. pointed out that a long random parametrized quantum circuit can resemble a Haar-random unitary map. Under such conditions, the typical value of the gradient of the cost function decreases exponentially with the number of qubits for any reasonable cost function. This finding is important for the design of any variational quantum algorithms.}

The proposed classification technique can be applied to any model that can be expressed as a spin model (e.g. fermion problems can be mapped to spin problems by using Jordan--Wigner transformation or Bravyi--Kitaev transformation).

\add{For the reported study, we tested three different ansatz circuits for VQE, but only one of them was used for classification. In principle, one could also try to use the other two ans\"atze for classification. Indeed, the rank-one ansatz would be a simple classifier that only uses local spin measurements. One could also use the tree tensor network, however it would have to be flipped compared to the VQE ansatz: the tree classifier would take $n$ qubits as input, and output only one or two qubits. However, as these two ans\"atze show poorer results for VQE, we do not believe they would admit improved results in their role as classification circuits. For a particular example, a rank-one classifier would work for the transverse field Ising model, because the alignment of spins along the X axis changes across the phase transition, as we mentioned before. For the XXZ model, however, this does not happen, and hence using the rank-one classifier would be pointless. Finally, the tree tensor network can be efficiently contracted on a classical machine, therefore the proposed classifier would not offer any computational advantage if it used the TTN as a classifier.}

We simulated quantum machine learning for binary classification. Such a protocol can be extended to more classes. For example, for four classes, we could treat ``0000000000'' as a label of class 1, ``0000011111'' as a label of class 2, ``1111100000'' as that of class 3, and finally ``1111111111'' as that of class 4. Other possibilities would then be labeled according to which of the four strings is closest in terms of Hamming distance.

\add{
    Since we have VQE solutions as the data points with assigned labels, $\{\ket{\psi(\boldsymbol{\theta}_i)}, y_i\}^{N_{train}}_{i=1}$, we could in principle use the $k$-nearest neighbors method \cite{hastie_friedman_tisbshirani_2017} for classifying a state $\ket{\rho}$.
    That is, for the simplest case with $k=1$, we would calculate all the $N_{train}$ overlaps $\mathcal{O}_i = |\langle \psi(\boldsymbol{\theta}_i)| \rho \rangle|^2$, find the the state $\ket{\psi(\boldsymbol{\theta}_m)}$ which gives the maximal overlap $\mathcal{O}_m$ and assign the label $y_m$ from the pair $\{\ket{\psi(\boldsymbol{\theta}_m)}, y_m\}$ to the state $\ket{\rho}$.
    If $\ket{\psi(\boldsymbol{\theta}_i)} = U(\boldsymbol{\theta}_i) \ket{\boldsymbol{0}}$ and $\ket{\rho} = V \ket{\boldsymbol{0}}$, then the overlap $\mathcal{O}_i$ can be calculated with a quantum computer by preparing the state $ V^\dagger U(\boldsymbol{\theta}_i)\ket{\boldsymbol{0}}$ and measuring it in the computational basis.
    The probability to obtain only zeros in the measurement result will be exactly the overlap $\mathcal{O}_i$.
    The main drawback of this approach is that in order to classify a state $\ket{\rho}$, one needs to estimate $N_{train}$ overlaps $\mathcal{O}$. In contrast, in the method we propose, once the classifier circuit $U_\mathrm{class}$ is trained, it requires only a single series of measurements to classify an input state. 
}

\begin{acknowledgments}
We thank Dmitry Yudin for fruitful discussions and suggestions. A.\,U. acknowledges RFBR project number 19-31-90159.

\paragraph*{Numerical Methods.}  The quantum circuits designed in the study were simulated using Qiskit \cite{aleksandrowicz_qiskit:_2019}. Figures \ref{fig:dE_Ising}, \ref{fig:classification} were prepared using Matplotlib \cite{hunter_matplotlib:_2007}. The source code for the study is available at \texttt{github.com/Quantum-Machine-Learning-Initiative/\\learning-phases}.
\end{acknowledgments}



\begin{thebibliography}{41}%
\makeatletter
\providecommand \@ifxundefined [1]{%
 \@ifx{#1\undefined}
}%
\providecommand \@ifnum [1]{%
 \ifnum #1\expandafter \@firstoftwo
 \else \expandafter \@secondoftwo
 \fi
}%
\providecommand \@ifx [1]{%
 \ifx #1\expandafter \@firstoftwo
 \else \expandafter \@secondoftwo
 \fi
}%
\providecommand \natexlab [1]{#1}%
\providecommand \enquote  [1]{``#1''}%
\providecommand \bibnamefont  [1]{#1}%
\providecommand \bibfnamefont [1]{#1}%
\providecommand \citenamefont [1]{#1}%
\providecommand \href@noop [0]{\@secondoftwo}%
\providecommand \href [0]{\begingroup \@sanitize@url \@href}%
\providecommand \@href[1]{\@@startlink{#1}\@@href}%
\providecommand \@@href[1]{\endgroup#1\@@endlink}%
\providecommand \@sanitize@url [0]{\catcode `\\12\catcode `\$12\catcode
  `\&12\catcode `\#12\catcode `\^12\catcode `\_12\catcode `\%12\relax}%
\providecommand \@@startlink[1]{}%
\providecommand \@@endlink[0]{}%
\providecommand \url  [0]{\begingroup\@sanitize@url \@url }%
\providecommand \@url [1]{\endgroup\@href {#1}{\urlprefix }}%
\providecommand \urlprefix  [0]{URL }%
\providecommand \Eprint [0]{\href }%
\providecommand \doibase [0]{http://dx.doi.org/}%
\providecommand \selectlanguage [0]{\@gobble}%
\providecommand \bibinfo  [0]{\@secondoftwo}%
\providecommand \bibfield  [0]{\@secondoftwo}%
\providecommand \translation [1]{[#1]}%
\providecommand \BibitemOpen [0]{}%
\providecommand \bibitemStop [0]{}%
\providecommand \bibitemNoStop [0]{.\EOS\space}%
\providecommand \EOS [0]{\spacefactor3000\relax}%
\providecommand \BibitemShut  [1]{\csname bibitem#1\endcsname}%
\let\auto@bib@innerbib\@empty
\bibitem [{\citenamefont {Carrasquilla}\ and\ \citenamefont
  {Melko}(2017)}]{carrasquilla_machine_2017}%
  \BibitemOpen
  \bibfield  {author} {\bibinfo {author} {\bibfnamefont {J.}~\bibnamefont
  {Carrasquilla}}\ and\ \bibinfo {author} {\bibfnamefont {R.~G.}\ \bibnamefont
  {Melko}},\ }\href {\doibase 10.1038/nphys4035} {\bibfield  {journal}
  {\bibinfo  {journal} {Nature Physics}\ }\textbf {\bibinfo {volume} {13}},\
  \bibinfo {pages} {431} (\bibinfo {year} {2017})},\ \bibinfo {note} {arXiv:
  1605.01735}\BibitemShut {NoStop}%
\bibitem [{\citenamefont {van Nieuwenburg}\ \emph {et~al.}(2017)\citenamefont
  {van Nieuwenburg}, \citenamefont {Liu},\ and\ \citenamefont
  {Huber}}]{van_nieuwenburg_learning_2017}%
  \BibitemOpen
  \bibfield  {author} {\bibinfo {author} {\bibfnamefont {E.}~\bibnamefont {van
  Nieuwenburg}}, \bibinfo {author} {\bibfnamefont {Y.-H.}\ \bibnamefont {Liu}},
  \ and\ \bibinfo {author} {\bibfnamefont {S.}~\bibnamefont {Huber}},\ }\href
  {\doibase 10.1038/nphys4037} {\bibfield  {journal} {\bibinfo  {journal}
  {Nature Physics}\ }\textbf {\bibinfo {volume} {13}},\ \bibinfo {pages} {435}
  (\bibinfo {year} {2017})}\BibitemShut {NoStop}%
\bibitem [{\citenamefont {Biamonte}\ \emph {et~al.}(2017)\citenamefont
  {Biamonte}, \citenamefont {Wittek}, \citenamefont {Pancotti}, \citenamefont
  {Rebentrost}, \citenamefont {Wiebe},\ and\ \citenamefont
  {Lloyd}}]{biamonte_quantum_2017}%
  \BibitemOpen
  \bibfield  {author} {\bibinfo {author} {\bibfnamefont {J.}~\bibnamefont
  {Biamonte}}, \bibinfo {author} {\bibfnamefont {P.}~\bibnamefont {Wittek}},
  \bibinfo {author} {\bibfnamefont {N.}~\bibnamefont {Pancotti}}, \bibinfo
  {author} {\bibfnamefont {P.}~\bibnamefont {Rebentrost}}, \bibinfo {author}
  {\bibfnamefont {N.}~\bibnamefont {Wiebe}}, \ and\ \bibinfo {author}
  {\bibfnamefont {S.}~\bibnamefont {Lloyd}},\ }\href {\doibase
  10.1038/nature23474} {\bibfield  {journal} {\bibinfo  {journal} {Nature}\
  }\textbf {\bibinfo {volume} {549}},\ \bibinfo {pages} {195} (\bibinfo {year}
  {2017})},\ \bibinfo {note} {arXiv: 1611.09347}\BibitemShut {NoStop}%
\bibitem [{\citenamefont {Huggins}\ \emph {et~al.}(2019)\citenamefont
  {Huggins}, \citenamefont {Patil}, \citenamefont {Mitchell}, \citenamefont
  {Whaley},\ and\ \citenamefont {Stoudenmire}}]{huggins_towards_2019}%
  \BibitemOpen
  \bibfield  {author} {\bibinfo {author} {\bibfnamefont {W.}~\bibnamefont
  {Huggins}}, \bibinfo {author} {\bibfnamefont {P.}~\bibnamefont {Patil}},
  \bibinfo {author} {\bibfnamefont {B.}~\bibnamefont {Mitchell}}, \bibinfo
  {author} {\bibfnamefont {K.~B.}\ \bibnamefont {Whaley}}, \ and\ \bibinfo
  {author} {\bibfnamefont {E.~M.}\ \bibnamefont {Stoudenmire}},\ }\href
  {\doibase 10.1088/2058-9565/aaea94} {\bibfield  {journal} {\bibinfo
  {journal} {Quantum Science and Technology}\ }\textbf {\bibinfo {volume}
  {4}},\ \bibinfo {pages} {024001} (\bibinfo {year} {2019})}\BibitemShut
  {NoStop}%
\bibitem [{\citenamefont {Schuld}\ \emph {et~al.}(2018)\citenamefont {Schuld},
  \citenamefont {Bocharov}, \citenamefont {Svore},\ and\ \citenamefont
  {Wiebe}}]{schuld_circuit-centric_2018}%
  \BibitemOpen
  \bibfield  {author} {\bibinfo {author} {\bibfnamefont {M.}~\bibnamefont
  {Schuld}}, \bibinfo {author} {\bibfnamefont {A.}~\bibnamefont {Bocharov}},
  \bibinfo {author} {\bibfnamefont {K.}~\bibnamefont {Svore}}, \ and\ \bibinfo
  {author} {\bibfnamefont {N.}~\bibnamefont {Wiebe}},\ }\href
  {http://arxiv.org/abs/1804.00633} {\bibfield  {journal} {\bibinfo  {journal}
  {arXiv:1804.00633 [quant-ph]}\ } (\bibinfo {year} {2018})},\ \bibinfo {note}
  {arXiv: 1804.00633}\BibitemShut {NoStop}%
\bibitem [{\citenamefont {Schuld}\ and\ \citenamefont
  {Killoran}(2019)}]{schuld_quantum_2019}%
  \BibitemOpen
  \bibfield  {author} {\bibinfo {author} {\bibfnamefont {M.}~\bibnamefont
  {Schuld}}\ and\ \bibinfo {author} {\bibfnamefont {N.}~\bibnamefont
  {Killoran}},\ }\href {\doibase 10.1103/PhysRevLett.122.040504} {\bibfield
  {journal} {\bibinfo  {journal} {Physical Review Letters}\ }\textbf {\bibinfo
  {volume} {122}},\ \bibinfo {pages} {040504} (\bibinfo {year} {2019})},\
  \bibinfo {note} {arXiv: 1803.07128}\BibitemShut {NoStop}%
\bibitem [{\citenamefont {Havlíček}\ \emph {et~al.}(2019)\citenamefont
  {Havlíček}, \citenamefont {Córcoles}, \citenamefont {Temme}, \citenamefont
  {Harrow}, \citenamefont {Kandala}, \citenamefont {Chow},\ and\ \citenamefont
  {Gambetta}}]{havlicek_supervised_2019}%
  \BibitemOpen
  \bibfield  {author} {\bibinfo {author} {\bibfnamefont {V.}~\bibnamefont
  {Havlíček}}, \bibinfo {author} {\bibfnamefont {A.~D.}\ \bibnamefont
  {Córcoles}}, \bibinfo {author} {\bibfnamefont {K.}~\bibnamefont {Temme}},
  \bibinfo {author} {\bibfnamefont {A.~W.}\ \bibnamefont {Harrow}}, \bibinfo
  {author} {\bibfnamefont {A.}~\bibnamefont {Kandala}}, \bibinfo {author}
  {\bibfnamefont {J.~M.}\ \bibnamefont {Chow}}, \ and\ \bibinfo {author}
  {\bibfnamefont {J.~M.}\ \bibnamefont {Gambetta}},\ }\href {\doibase
  10.1038/s41586-019-0980-2} {\bibfield  {journal} {\bibinfo  {journal}
  {Nature}\ }\textbf {\bibinfo {volume} {567}},\ \bibinfo {pages} {209}
  (\bibinfo {year} {2019})}\BibitemShut {NoStop}%
\bibitem [{\citenamefont {Schuld}\ \emph {et~al.}(2015)\citenamefont {Schuld},
  \citenamefont {Sinayskiy},\ and\ \citenamefont
  {Petruccione}}]{schuld_introduction_2015}%
  \BibitemOpen
  \bibfield  {author} {\bibinfo {author} {\bibfnamefont {M.}~\bibnamefont
  {Schuld}}, \bibinfo {author} {\bibfnamefont {I.}~\bibnamefont {Sinayskiy}}, \
  and\ \bibinfo {author} {\bibfnamefont {F.}~\bibnamefont {Petruccione}},\
  }\href {\doibase 10.1080/00107514.2014.964942} {\bibfield  {journal}
  {\bibinfo  {journal} {Contemporary Physics}\ }\textbf {\bibinfo {volume}
  {56}},\ \bibinfo {pages} {172} (\bibinfo {year} {2015})},\ \bibinfo {note}
  {arXiv: 1409.3097}\BibitemShut {NoStop}%
\bibitem [{\citenamefont {Duan}\ \emph {et~al.}(2017)\citenamefont {Duan},
  \citenamefont {Yuan}, \citenamefont {Liu},\ and\ \citenamefont
  {Li}}]{duan_quantum_2017}%
  \BibitemOpen
  \bibfield  {author} {\bibinfo {author} {\bibfnamefont {B.}~\bibnamefont
  {Duan}}, \bibinfo {author} {\bibfnamefont {J.}~\bibnamefont {Yuan}}, \bibinfo
  {author} {\bibfnamefont {Y.}~\bibnamefont {Liu}}, \ and\ \bibinfo {author}
  {\bibfnamefont {D.}~\bibnamefont {Li}},\ }\href {\doibase
  10.1103/PhysRevA.96.032301} {\bibfield  {journal} {\bibinfo  {journal}
  {Physical Review A}\ }\textbf {\bibinfo {volume} {96}},\ \bibinfo {pages}
  {032301} (\bibinfo {year} {2017})}\BibitemShut {NoStop}%
\bibitem [{\citenamefont {Duan}\ \emph {et~al.}(2019)\citenamefont {Duan},
  \citenamefont {Yuan}, \citenamefont {Xu},\ and\ \citenamefont
  {Li}}]{duan_a_optimal_2019}%
  \BibitemOpen
  \bibfield  {author} {\bibinfo {author} {\bibfnamefont {B.}~\bibnamefont
  {Duan}}, \bibinfo {author} {\bibfnamefont {J.}~\bibnamefont {Yuan}}, \bibinfo
  {author} {\bibfnamefont {J.}~\bibnamefont {Xu}}, \ and\ \bibinfo {author}
  {\bibfnamefont {D.}~\bibnamefont {Li}},\ }\href {\doibase
  10.1103/PhysRevA.99.032311} {\bibfield  {journal} {\bibinfo  {journal}
  {Physical Review A}\ }\textbf {\bibinfo {volume} {99}},\ \bibinfo {pages}
  {032311} (\bibinfo {year} {2019})}\BibitemShut {NoStop}%
\bibitem [{\citenamefont {Sheng}\ and\ \citenamefont
  {Zhou}(2017)}]{sheng_distributed_2017}%
  \BibitemOpen
  \bibfield  {author} {\bibinfo {author} {\bibfnamefont {Y.-B.}\ \bibnamefont
  {Sheng}}\ and\ \bibinfo {author} {\bibfnamefont {L.}~\bibnamefont {Zhou}},\
  }\href {\doibase 10.1016/j.scib.2017.06.007} {\bibfield  {journal} {\bibinfo
  {journal} {Science Bulletin}\ }\textbf {\bibinfo {volume} {62}},\ \bibinfo
  {pages} {1025–1029} (\bibinfo {year} {2017})}\BibitemShut {NoStop}%
\bibitem [{\citenamefont {Georgescu}\ \emph {et~al.}(2014)\citenamefont
  {Georgescu}, \citenamefont {Ashhab},\ and\ \citenamefont
  {Nori}}]{georgescu_quantum_2014}%
  \BibitemOpen
  \bibfield  {author} {\bibinfo {author} {\bibfnamefont {I.}~\bibnamefont
  {Georgescu}}, \bibinfo {author} {\bibfnamefont {S.}~\bibnamefont {Ashhab}}, \
  and\ \bibinfo {author} {\bibfnamefont {F.}~\bibnamefont {Nori}},\ }\href
  {\doibase 10.1103/RevModPhys.86.153} {\bibfield  {journal} {\bibinfo
  {journal} {Reviews of Modern Physics}\ }\textbf {\bibinfo {volume} {86}},\
  \bibinfo {pages} {153–185} (\bibinfo {year} {2014})}\BibitemShut {NoStop}%
\bibitem [{\citenamefont {Bernien}\ \emph {et~al.}(2017)\citenamefont
  {Bernien}, \citenamefont {Schwartz}, \citenamefont {Keesling}, \citenamefont
  {Levine}, \citenamefont {Omran}, \citenamefont {Pichler}, \citenamefont
  {Choi}, \citenamefont {Zibrov}, \citenamefont {Endres}, \citenamefont
  {Greiner}, \citenamefont {Vuletić},\ and\ \citenamefont
  {Lukin}}]{bernien_probing_2017}%
  \BibitemOpen
  \bibfield  {author} {\bibinfo {author} {\bibfnamefont {H.}~\bibnamefont
  {Bernien}}, \bibinfo {author} {\bibfnamefont {S.}~\bibnamefont {Schwartz}},
  \bibinfo {author} {\bibfnamefont {A.}~\bibnamefont {Keesling}}, \bibinfo
  {author} {\bibfnamefont {H.}~\bibnamefont {Levine}}, \bibinfo {author}
  {\bibfnamefont {A.}~\bibnamefont {Omran}}, \bibinfo {author} {\bibfnamefont
  {H.}~\bibnamefont {Pichler}}, \bibinfo {author} {\bibfnamefont
  {S.}~\bibnamefont {Choi}}, \bibinfo {author} {\bibfnamefont {A.~S.}\
  \bibnamefont {Zibrov}}, \bibinfo {author} {\bibfnamefont {M.}~\bibnamefont
  {Endres}}, \bibinfo {author} {\bibfnamefont {M.}~\bibnamefont {Greiner}},
  \bibinfo {author} {\bibfnamefont {V.}~\bibnamefont {Vuletić}}, \ and\
  \bibinfo {author} {\bibfnamefont {M.~D.}\ \bibnamefont {Lukin}},\ }\href
  {\doibase 10.1038/nature24622} {\bibfield  {journal} {\bibinfo  {journal}
  {Nature}\ }\textbf {\bibinfo {volume} {551}},\ \bibinfo {pages} {579}
  (\bibinfo {year} {2017})}\BibitemShut {NoStop}%
\bibitem [{\citenamefont {Barreiro}\ \emph {et~al.}(2011)\citenamefont
  {Barreiro}, \citenamefont {Müller}, \citenamefont {Schindler}, \citenamefont
  {Nigg}, \citenamefont {Monz}, \citenamefont {Chwalla}, \citenamefont
  {Hennrich}, \citenamefont {Roos}, \citenamefont {Zoller},\ and\ \citenamefont
  {Blatt}}]{barreiro_open-system_2011}%
  \BibitemOpen
  \bibfield  {author} {\bibinfo {author} {\bibfnamefont {J.~T.}\ \bibnamefont
  {Barreiro}}, \bibinfo {author} {\bibfnamefont {M.}~\bibnamefont {Müller}},
  \bibinfo {author} {\bibfnamefont {P.}~\bibnamefont {Schindler}}, \bibinfo
  {author} {\bibfnamefont {D.}~\bibnamefont {Nigg}}, \bibinfo {author}
  {\bibfnamefont {T.}~\bibnamefont {Monz}}, \bibinfo {author} {\bibfnamefont
  {M.}~\bibnamefont {Chwalla}}, \bibinfo {author} {\bibfnamefont
  {M.}~\bibnamefont {Hennrich}}, \bibinfo {author} {\bibfnamefont {C.~F.}\
  \bibnamefont {Roos}}, \bibinfo {author} {\bibfnamefont {P.}~\bibnamefont
  {Zoller}}, \ and\ \bibinfo {author} {\bibfnamefont {R.}~\bibnamefont
  {Blatt}},\ }\href {\doibase 10.1038/nature09801} {\bibfield  {journal}
  {\bibinfo  {journal} {Nature}\ }\textbf {\bibinfo {volume} {470}},\ \bibinfo
  {pages} {486–491} (\bibinfo {year} {2011})}\BibitemShut {NoStop}%
\bibitem [{\citenamefont {Peruzzo}\ \emph {et~al.}(2014)\citenamefont
  {Peruzzo}, \citenamefont {McClean}, \citenamefont {Shadbolt}, \citenamefont
  {Yung}, \citenamefont {Zhou}, \citenamefont {Love}, \citenamefont
  {Aspuru-Guzik},\ and\ \citenamefont {O’Brien}}]{peruzzo_variational_2014}%
  \BibitemOpen
  \bibfield  {author} {\bibinfo {author} {\bibfnamefont {A.}~\bibnamefont
  {Peruzzo}}, \bibinfo {author} {\bibfnamefont {J.}~\bibnamefont {McClean}},
  \bibinfo {author} {\bibfnamefont {P.}~\bibnamefont {Shadbolt}}, \bibinfo
  {author} {\bibfnamefont {M.-H.}\ \bibnamefont {Yung}}, \bibinfo {author}
  {\bibfnamefont {X.-Q.}\ \bibnamefont {Zhou}}, \bibinfo {author}
  {\bibfnamefont {P.~J.}\ \bibnamefont {Love}}, \bibinfo {author}
  {\bibfnamefont {A.}~\bibnamefont {Aspuru-Guzik}}, \ and\ \bibinfo {author}
  {\bibfnamefont {J.~L.}\ \bibnamefont {O’Brien}},\ }\href {\doibase
  10.1038/ncomms5213} {\bibfield  {journal} {\bibinfo  {journal} {Nature
  Communications}\ }\textbf {\bibinfo {volume} {5}} (\bibinfo {year} {2014}),\
  10.1038/ncomms5213}\BibitemShut {NoStop}%
\bibitem [{\citenamefont {Wang}(2016)}]{wang_discovering_2016}%
  \BibitemOpen
  \bibfield  {author} {\bibinfo {author} {\bibfnamefont {L.}~\bibnamefont
  {Wang}},\ }\href {\doibase 10.1103/PhysRevB.94.195105} {\bibfield  {journal}
  {\bibinfo  {journal} {Physical Review B}\ }\textbf {\bibinfo {volume} {94}},\
  \bibinfo {pages} {195105} (\bibinfo {year} {2016})}\BibitemShut {NoStop}%
\bibitem [{\citenamefont {McClean}\ \emph {et~al.}(2016)\citenamefont
  {McClean}, \citenamefont {Romero}, \citenamefont {Babbush},\ and\
  \citenamefont {Aspuru-Guzik}}]{mcclean_theory_2016}%
  \BibitemOpen
  \bibfield  {author} {\bibinfo {author} {\bibfnamefont {J.~R.}\ \bibnamefont
  {McClean}}, \bibinfo {author} {\bibfnamefont {J.}~\bibnamefont {Romero}},
  \bibinfo {author} {\bibfnamefont {R.}~\bibnamefont {Babbush}}, \ and\
  \bibinfo {author} {\bibfnamefont {A.}~\bibnamefont {Aspuru-Guzik}},\ }\href
  {\doibase 10.1088/1367-2630/18/2/023023} {\bibfield  {journal} {\bibinfo
  {journal} {New Journal of Physics}\ }\textbf {\bibinfo {volume} {18}},\
  \bibinfo {pages} {023023} (\bibinfo {year} {2016})},\ \bibinfo {note} {arXiv:
  1509.04279}\BibitemShut {NoStop}%
\bibitem [{\citenamefont {Wecker}\ \emph {et~al.}(2015)\citenamefont {Wecker},
  \citenamefont {Hastings},\ and\ \citenamefont
  {Troyer}}]{wecker_progress_2015}%
  \BibitemOpen
  \bibfield  {author} {\bibinfo {author} {\bibfnamefont {D.}~\bibnamefont
  {Wecker}}, \bibinfo {author} {\bibfnamefont {M.~B.}\ \bibnamefont
  {Hastings}}, \ and\ \bibinfo {author} {\bibfnamefont {M.}~\bibnamefont
  {Troyer}},\ }\href {\doibase 10.1103/PhysRevA.92.042303} {\bibfield
  {journal} {\bibinfo  {journal} {Physical Review A}\ }\textbf {\bibinfo
  {volume} {92}},\ \bibinfo {pages} {042303} (\bibinfo {year}
  {2015})}\BibitemShut {NoStop}%
\bibitem [{\citenamefont {Khatri}\ \emph {et~al.}(2018)\citenamefont {Khatri},
  \citenamefont {LaRose}, \citenamefont {Poremba}, \citenamefont {Cincio},
  \citenamefont {Sornborger},\ and\ \citenamefont
  {Coles}}]{khatri_quantum-assisted_2018}%
  \BibitemOpen
  \bibfield  {author} {\bibinfo {author} {\bibfnamefont {S.}~\bibnamefont
  {Khatri}}, \bibinfo {author} {\bibfnamefont {R.}~\bibnamefont {LaRose}},
  \bibinfo {author} {\bibfnamefont {A.}~\bibnamefont {Poremba}}, \bibinfo
  {author} {\bibfnamefont {L.}~\bibnamefont {Cincio}}, \bibinfo {author}
  {\bibfnamefont {A.~T.}\ \bibnamefont {Sornborger}}, \ and\ \bibinfo {author}
  {\bibfnamefont {P.~J.}\ \bibnamefont {Coles}},\ }\href
  {http://arxiv.org/abs/1807.00800} {\bibfield  {journal} {\bibinfo  {journal}
  {arXiv:1807.00800 [quant-ph]}\ } (\bibinfo {year} {2018})},\ \bibinfo {note}
  {arXiv: 1807.00800}\BibitemShut {NoStop}%
\bibitem [{\citenamefont {O'Malley}\ \emph {et~al.}(2016)\citenamefont
  {O'Malley}, \citenamefont {Babbush}, \citenamefont {Kivlichan}, \citenamefont
  {Romero}, \citenamefont {McClean}, \citenamefont {Barends}, \citenamefont
  {Kelly}, \citenamefont {Roushan}, \citenamefont {Tranter}, \citenamefont
  {Ding}, \citenamefont {Campbell}, \citenamefont {Chen}, \citenamefont {Chen},
  \citenamefont {Chiaro}, \citenamefont {Dunsworth}, \citenamefont {Fowler},
  \citenamefont {Jeffrey}, \citenamefont {Megrant}, \citenamefont {Mutus},
  \citenamefont {Neill}, \citenamefont {Quintana}, \citenamefont {Sank},
  \citenamefont {Vainsencher}, \citenamefont {Wenner}, \citenamefont {White},
  \citenamefont {Coveney}, \citenamefont {Love}, \citenamefont {Neven},
  \citenamefont {Aspuru-Guzik},\ and\ \citenamefont
  {Martinis}}]{omalley_scalable_2016}%
  \BibitemOpen
  \bibfield  {author} {\bibinfo {author} {\bibfnamefont {P.~J.~J.}\
  \bibnamefont {O'Malley}}, \bibinfo {author} {\bibfnamefont {R.}~\bibnamefont
  {Babbush}}, \bibinfo {author} {\bibfnamefont {I.~D.}\ \bibnamefont
  {Kivlichan}}, \bibinfo {author} {\bibfnamefont {J.}~\bibnamefont {Romero}},
  \bibinfo {author} {\bibfnamefont {J.~R.}\ \bibnamefont {McClean}}, \bibinfo
  {author} {\bibfnamefont {R.}~\bibnamefont {Barends}}, \bibinfo {author}
  {\bibfnamefont {J.}~\bibnamefont {Kelly}}, \bibinfo {author} {\bibfnamefont
  {P.}~\bibnamefont {Roushan}}, \bibinfo {author} {\bibfnamefont
  {A.}~\bibnamefont {Tranter}}, \bibinfo {author} {\bibfnamefont
  {N.}~\bibnamefont {Ding}}, \bibinfo {author} {\bibfnamefont {B.}~\bibnamefont
  {Campbell}}, \bibinfo {author} {\bibfnamefont {Y.}~\bibnamefont {Chen}},
  \bibinfo {author} {\bibfnamefont {Z.}~\bibnamefont {Chen}}, \bibinfo {author}
  {\bibfnamefont {B.}~\bibnamefont {Chiaro}}, \bibinfo {author} {\bibfnamefont
  {A.}~\bibnamefont {Dunsworth}}, \bibinfo {author} {\bibfnamefont {A.~G.}\
  \bibnamefont {Fowler}}, \bibinfo {author} {\bibfnamefont {E.}~\bibnamefont
  {Jeffrey}}, \bibinfo {author} {\bibfnamefont {A.}~\bibnamefont {Megrant}},
  \bibinfo {author} {\bibfnamefont {J.~Y.}\ \bibnamefont {Mutus}}, \bibinfo
  {author} {\bibfnamefont {C.}~\bibnamefont {Neill}}, \bibinfo {author}
  {\bibfnamefont {C.}~\bibnamefont {Quintana}}, \bibinfo {author}
  {\bibfnamefont {D.}~\bibnamefont {Sank}}, \bibinfo {author} {\bibfnamefont
  {A.}~\bibnamefont {Vainsencher}}, \bibinfo {author} {\bibfnamefont
  {J.}~\bibnamefont {Wenner}}, \bibinfo {author} {\bibfnamefont {T.~C.}\
  \bibnamefont {White}}, \bibinfo {author} {\bibfnamefont {P.~V.}\ \bibnamefont
  {Coveney}}, \bibinfo {author} {\bibfnamefont {P.~J.}\ \bibnamefont {Love}},
  \bibinfo {author} {\bibfnamefont {H.}~\bibnamefont {Neven}}, \bibinfo
  {author} {\bibfnamefont {A.}~\bibnamefont {Aspuru-Guzik}}, \ and\ \bibinfo
  {author} {\bibfnamefont {J.~M.}\ \bibnamefont {Martinis}},\ }\href {\doibase
  10.1103/PhysRevX.6.031007} {\bibfield  {journal} {\bibinfo  {journal}
  {Physical Review X}\ }\textbf {\bibinfo {volume} {6}} (\bibinfo {year}
  {2016}),\ 10.1103/PhysRevX.6.031007},\ \bibinfo {note} {arXiv:
  1512.06860}\BibitemShut {NoStop}%
\bibitem [{\citenamefont {Colless}\ \emph {et~al.}(2018)\citenamefont
  {Colless}, \citenamefont {Ramasesh}, \citenamefont {Dahlen}, \citenamefont
  {Blok}, \citenamefont {Kimchi-Schwartz}, \citenamefont {McClean},
  \citenamefont {Carter}, \citenamefont {de~Jong},\ and\ \citenamefont
  {Siddiqi}}]{colless_computation_2018}%
  \BibitemOpen
  \bibfield  {author} {\bibinfo {author} {\bibfnamefont {J.}~\bibnamefont
  {Colless}}, \bibinfo {author} {\bibfnamefont {V.}~\bibnamefont {Ramasesh}},
  \bibinfo {author} {\bibfnamefont {D.}~\bibnamefont {Dahlen}}, \bibinfo
  {author} {\bibfnamefont {M.}~\bibnamefont {Blok}}, \bibinfo {author}
  {\bibfnamefont {M.}~\bibnamefont {Kimchi-Schwartz}}, \bibinfo {author}
  {\bibfnamefont {J.}~\bibnamefont {McClean}}, \bibinfo {author} {\bibfnamefont
  {J.}~\bibnamefont {Carter}}, \bibinfo {author} {\bibfnamefont
  {W.}~\bibnamefont {de~Jong}}, \ and\ \bibinfo {author} {\bibfnamefont
  {I.}~\bibnamefont {Siddiqi}},\ }\href {\doibase 10.1103/PhysRevX.8.011021}
  {\bibfield  {journal} {\bibinfo  {journal} {Physical Review X}\ }\textbf
  {\bibinfo {volume} {8}} (\bibinfo {year} {2018}),\
  10.1103/PhysRevX.8.011021}\BibitemShut {NoStop}%
\bibitem [{\citenamefont {Shen}\ \emph {et~al.}(2017)\citenamefont {Shen},
  \citenamefont {Zhang}, \citenamefont {Zhang}, \citenamefont {Zhang},
  \citenamefont {Yung},\ and\ \citenamefont {Kim}}]{shen_quantum_2017}%
  \BibitemOpen
  \bibfield  {author} {\bibinfo {author} {\bibfnamefont {Y.}~\bibnamefont
  {Shen}}, \bibinfo {author} {\bibfnamefont {X.}~\bibnamefont {Zhang}},
  \bibinfo {author} {\bibfnamefont {S.}~\bibnamefont {Zhang}}, \bibinfo
  {author} {\bibfnamefont {J.-N.}\ \bibnamefont {Zhang}}, \bibinfo {author}
  {\bibfnamefont {M.-H.}\ \bibnamefont {Yung}}, \ and\ \bibinfo {author}
  {\bibfnamefont {K.}~\bibnamefont {Kim}},\ }\href {\doibase
  10.1103/PhysRevA.95.020501} {\bibfield  {journal} {\bibinfo  {journal}
  {Physical Review A}\ }\textbf {\bibinfo {volume} {95}} (\bibinfo {year}
  {2017}),\ 10.1103/PhysRevA.95.020501},\ \bibinfo {note} {arXiv:
  1506.00443}\BibitemShut {NoStop}%
\bibitem [{\citenamefont {Hempel}\ \emph {et~al.}(2018)\citenamefont {Hempel},
  \citenamefont {Maier}, \citenamefont {Romero}, \citenamefont {McClean},
  \citenamefont {Monz}, \citenamefont {Shen}, \citenamefont {Jurcevic},
  \citenamefont {Lanyon}, \citenamefont {Love}, \citenamefont {Babbush},
  \citenamefont {Aspuru-Guzik}, \citenamefont {Blatt},\ and\ \citenamefont
  {Roos}}]{hempel_quantum_2018}%
  \BibitemOpen
  \bibfield  {author} {\bibinfo {author} {\bibfnamefont {C.}~\bibnamefont
  {Hempel}}, \bibinfo {author} {\bibfnamefont {C.}~\bibnamefont {Maier}},
  \bibinfo {author} {\bibfnamefont {J.}~\bibnamefont {Romero}}, \bibinfo
  {author} {\bibfnamefont {J.}~\bibnamefont {McClean}}, \bibinfo {author}
  {\bibfnamefont {T.}~\bibnamefont {Monz}}, \bibinfo {author} {\bibfnamefont
  {H.}~\bibnamefont {Shen}}, \bibinfo {author} {\bibfnamefont {P.}~\bibnamefont
  {Jurcevic}}, \bibinfo {author} {\bibfnamefont {B.~P.}\ \bibnamefont
  {Lanyon}}, \bibinfo {author} {\bibfnamefont {P.}~\bibnamefont {Love}},
  \bibinfo {author} {\bibfnamefont {R.}~\bibnamefont {Babbush}}, \bibinfo
  {author} {\bibfnamefont {A.}~\bibnamefont {Aspuru-Guzik}}, \bibinfo {author}
  {\bibfnamefont {R.}~\bibnamefont {Blatt}}, \ and\ \bibinfo {author}
  {\bibfnamefont {C.~F.}\ \bibnamefont {Roos}},\ }\href {\doibase
  10.1103/PhysRevX.8.031022} {\bibfield  {journal} {\bibinfo  {journal}
  {Physical Review X}\ }\textbf {\bibinfo {volume} {8}} (\bibinfo {year}
  {2018}),\ 10.1103/PhysRevX.8.031022}\BibitemShut {NoStop}%
\bibitem [{\citenamefont {Dumitrescu}\ \emph {et~al.}(2018)\citenamefont
  {Dumitrescu}, \citenamefont {McCaskey}, \citenamefont {Hagen}, \citenamefont
  {Jansen}, \citenamefont {Morris}, \citenamefont {Papenbrock}, \citenamefont
  {Pooser}, \citenamefont {Dean},\ and\ \citenamefont
  {Lougovski}}]{dumitrescu_cloud_2018}%
  \BibitemOpen
  \bibfield  {author} {\bibinfo {author} {\bibfnamefont {E.~F.}\ \bibnamefont
  {Dumitrescu}}, \bibinfo {author} {\bibfnamefont {A.~J.}\ \bibnamefont
  {McCaskey}}, \bibinfo {author} {\bibfnamefont {G.}~\bibnamefont {Hagen}},
  \bibinfo {author} {\bibfnamefont {G.~R.}\ \bibnamefont {Jansen}}, \bibinfo
  {author} {\bibfnamefont {T.~D.}\ \bibnamefont {Morris}}, \bibinfo {author}
  {\bibfnamefont {T.}~\bibnamefont {Papenbrock}}, \bibinfo {author}
  {\bibfnamefont {R.~C.}\ \bibnamefont {Pooser}}, \bibinfo {author}
  {\bibfnamefont {D.~J.}\ \bibnamefont {Dean}}, \ and\ \bibinfo {author}
  {\bibfnamefont {P.}~\bibnamefont {Lougovski}},\ }\href {\doibase
  10.1103/PhysRevLett.120.210501} {\bibfield  {journal} {\bibinfo  {journal}
  {Physical Review Letters}\ }\textbf {\bibinfo {volume} {120}} (\bibinfo
  {year} {2018}),\ 10.1103/PhysRevLett.120.210501},\ \bibinfo {note} {arXiv:
  1801.03897}\BibitemShut {NoStop}%
\bibitem [{\citenamefont {Götze}\ \emph {et~al.}(2011)\citenamefont {Götze},
  \citenamefont {Farnell}, \citenamefont {Bishop}, \citenamefont {Li},\ and\
  \citenamefont {Richter}}]{gotze_heisenberg_2011}%
  \BibitemOpen
  \bibfield  {author} {\bibinfo {author} {\bibfnamefont {O.}~\bibnamefont
  {Götze}}, \bibinfo {author} {\bibfnamefont {D.~J.~J.}\ \bibnamefont
  {Farnell}}, \bibinfo {author} {\bibfnamefont {R.~F.}\ \bibnamefont {Bishop}},
  \bibinfo {author} {\bibfnamefont {P.~H.~Y.}\ \bibnamefont {Li}}, \ and\
  \bibinfo {author} {\bibfnamefont {J.}~\bibnamefont {Richter}},\ }\href
  {\doibase 10.1103/PhysRevB.84.224428} {\bibfield  {journal} {\bibinfo
  {journal} {Physical Review B}\ }\textbf {\bibinfo {volume} {84}} (\bibinfo
  {year} {2011}),\ 10.1103/PhysRevB.84.224428}\BibitemShut {NoStop}%
\bibitem [{\citenamefont {Taube}\ and\ \citenamefont
  {Bartlett}(2006)}]{taube_new_2006}%
  \BibitemOpen
  \bibfield  {author} {\bibinfo {author} {\bibfnamefont {A.~G.}\ \bibnamefont
  {Taube}}\ and\ \bibinfo {author} {\bibfnamefont {R.~J.}\ \bibnamefont
  {Bartlett}},\ }\href {\doibase 10.1002/qua.21198} {\bibfield  {journal}
  {\bibinfo  {journal} {International Journal of Quantum Chemistry}\ }\textbf
  {\bibinfo {volume} {106}},\ \bibinfo {pages} {3393} (\bibinfo {year}
  {2006})}\BibitemShut {NoStop}%
\bibitem [{\citenamefont {Lanyon}\ \emph {et~al.}(2010)\citenamefont {Lanyon},
  \citenamefont {Whitfield}, \citenamefont {Gillet}, \citenamefont {Goggin},
  \citenamefont {Almeida}, \citenamefont {Kassal}, \citenamefont {Biamonte},
  \citenamefont {Mohseni}, \citenamefont {Powell}, \citenamefont {Barbieri},
  \citenamefont {Aspuru-Guzik},\ and\ \citenamefont
  {White}}]{lanyon_towards_2010}%
  \BibitemOpen
  \bibfield  {author} {\bibinfo {author} {\bibfnamefont {B.~P.}\ \bibnamefont
  {Lanyon}}, \bibinfo {author} {\bibfnamefont {J.~D.}\ \bibnamefont
  {Whitfield}}, \bibinfo {author} {\bibfnamefont {G.~G.}\ \bibnamefont
  {Gillet}}, \bibinfo {author} {\bibfnamefont {M.~E.}\ \bibnamefont {Goggin}},
  \bibinfo {author} {\bibfnamefont {M.~P.}\ \bibnamefont {Almeida}}, \bibinfo
  {author} {\bibfnamefont {I.}~\bibnamefont {Kassal}}, \bibinfo {author}
  {\bibfnamefont {J.~D.}\ \bibnamefont {Biamonte}}, \bibinfo {author}
  {\bibfnamefont {M.}~\bibnamefont {Mohseni}}, \bibinfo {author} {\bibfnamefont
  {B.~J.}\ \bibnamefont {Powell}}, \bibinfo {author} {\bibfnamefont
  {M.}~\bibnamefont {Barbieri}}, \bibinfo {author} {\bibfnamefont
  {A.}~\bibnamefont {Aspuru-Guzik}}, \ and\ \bibinfo {author} {\bibfnamefont
  {A.~G.}\ \bibnamefont {White}},\ }\href {\doibase 10.1038/nchem.483}
  {\bibfield  {journal} {\bibinfo  {journal} {Nature Chemistry}\ }\textbf
  {\bibinfo {volume} {2}},\ \bibinfo {pages} {106} (\bibinfo {year} {2010})},\
  \bibinfo {note} {arXiv: 0905.0887}\BibitemShut {NoStop}%
\bibitem [{\citenamefont {Cao}\ \emph {et~al.}(2018)\citenamefont {Cao},
  \citenamefont {Romero}, \citenamefont {Olson}, \citenamefont {Degroote},
  \citenamefont {Johnson}, \citenamefont {Kieferová}, \citenamefont
  {Kivlichan}, \citenamefont {Menke}, \citenamefont {Peropadre}, \citenamefont
  {Sawaya}, \citenamefont {Sim}, \citenamefont {Veis},\ and\ \citenamefont
  {Aspuru-Guzik}}]{cao_quantum_2018}%
  \BibitemOpen
  \bibfield  {author} {\bibinfo {author} {\bibfnamefont {Y.}~\bibnamefont
  {Cao}}, \bibinfo {author} {\bibfnamefont {J.}~\bibnamefont {Romero}},
  \bibinfo {author} {\bibfnamefont {J.~P.}\ \bibnamefont {Olson}}, \bibinfo
  {author} {\bibfnamefont {M.}~\bibnamefont {Degroote}}, \bibinfo {author}
  {\bibfnamefont {P.~D.}\ \bibnamefont {Johnson}}, \bibinfo {author}
  {\bibfnamefont {M.}~\bibnamefont {Kieferová}}, \bibinfo {author}
  {\bibfnamefont {I.~D.}\ \bibnamefont {Kivlichan}}, \bibinfo {author}
  {\bibfnamefont {T.}~\bibnamefont {Menke}}, \bibinfo {author} {\bibfnamefont
  {B.}~\bibnamefont {Peropadre}}, \bibinfo {author} {\bibfnamefont {N.~P.~D.}\
  \bibnamefont {Sawaya}}, \bibinfo {author} {\bibfnamefont {S.}~\bibnamefont
  {Sim}}, \bibinfo {author} {\bibfnamefont {L.}~\bibnamefont {Veis}}, \ and\
  \bibinfo {author} {\bibfnamefont {A.}~\bibnamefont {Aspuru-Guzik}},\ }\href
  {http://arxiv.org/abs/1812.09976} {\bibfield  {journal} {\bibinfo  {journal}
  {arXiv:1812.09976 [quant-ph]}\ } (\bibinfo {year} {2018})},\ \bibinfo {note}
  {arXiv: 1812.09976}\BibitemShut {NoStop}%
\bibitem [{\citenamefont {Vatan}\ and\ \citenamefont
  {Williams}(2004)}]{vatan_optimal_2004}%
  \BibitemOpen
  \bibfield  {author} {\bibinfo {author} {\bibfnamefont {F.}~\bibnamefont
  {Vatan}}\ and\ \bibinfo {author} {\bibfnamefont {C.}~\bibnamefont
  {Williams}},\ }\href {\doibase 10.1103/PhysRevA.69.032315} {\bibfield
  {journal} {\bibinfo  {journal} {Physical Review A}\ }\textbf {\bibinfo
  {volume} {69}} (\bibinfo {year} {2004}),\ 10.1103/PhysRevA.69.032315},\
  \bibinfo {note} {arXiv: quant-ph/0308006}\BibitemShut {NoStop}%
\bibitem [{\citenamefont {Santagati}\ \emph {et~al.}(2018)\citenamefont
  {Santagati}, \citenamefont {Wang}, \citenamefont {Gentile}, \citenamefont
  {Paesani}, \citenamefont {Wiebe}, \citenamefont {McClean}, \citenamefont
  {Morley-Short}, \citenamefont {Shadbolt}, \citenamefont {Bonneau},
  \citenamefont {Silverstone}, \citenamefont {Tew}, \citenamefont {Zhou},
  \citenamefont {O’Brien},\ and\ \citenamefont
  {Thompson}}]{santagati_witnessing_2018}%
  \BibitemOpen
  \bibfield  {author} {\bibinfo {author} {\bibfnamefont {R.}~\bibnamefont
  {Santagati}}, \bibinfo {author} {\bibfnamefont {J.}~\bibnamefont {Wang}},
  \bibinfo {author} {\bibfnamefont {A.~A.}\ \bibnamefont {Gentile}}, \bibinfo
  {author} {\bibfnamefont {S.}~\bibnamefont {Paesani}}, \bibinfo {author}
  {\bibfnamefont {N.}~\bibnamefont {Wiebe}}, \bibinfo {author} {\bibfnamefont
  {J.~R.}\ \bibnamefont {McClean}}, \bibinfo {author} {\bibfnamefont
  {S.}~\bibnamefont {Morley-Short}}, \bibinfo {author} {\bibfnamefont {P.~J.}\
  \bibnamefont {Shadbolt}}, \bibinfo {author} {\bibfnamefont {D.}~\bibnamefont
  {Bonneau}}, \bibinfo {author} {\bibfnamefont {J.~W.}\ \bibnamefont
  {Silverstone}}, \bibinfo {author} {\bibfnamefont {D.~P.}\ \bibnamefont
  {Tew}}, \bibinfo {author} {\bibfnamefont {X.}~\bibnamefont {Zhou}}, \bibinfo
  {author} {\bibfnamefont {J.~L.}\ \bibnamefont {O’Brien}}, \ and\ \bibinfo
  {author} {\bibfnamefont {M.~G.}\ \bibnamefont {Thompson}},\ }\href {\doibase
  10.1126/sciadv.aap9646} {\bibfield  {journal} {\bibinfo  {journal} {Science
  Advances}\ }\textbf {\bibinfo {volume} {4}},\ \bibinfo {pages} {eaap9646}
  (\bibinfo {year} {2018})}\BibitemShut {NoStop}%
\bibitem [{\citenamefont {Farhi}\ \emph {et~al.}(2014)\citenamefont {Farhi},
  \citenamefont {Goldstone},\ and\ \citenamefont
  {Gutmann}}]{farhi_quantum_2014}%
  \BibitemOpen
  \bibfield  {author} {\bibinfo {author} {\bibfnamefont {E.}~\bibnamefont
  {Farhi}}, \bibinfo {author} {\bibfnamefont {J.}~\bibnamefont {Goldstone}}, \
  and\ \bibinfo {author} {\bibfnamefont {S.}~\bibnamefont {Gutmann}},\ }\href
  {http://arxiv.org/abs/1411.4028} {\bibfield  {journal} {\bibinfo  {journal}
  {arXiv:1411.4028 [quant-ph]}\ } (\bibinfo {year} {2014})},\ \bibinfo {note}
  {arXiv: 1411.4028}\BibitemShut {NoStop}%
\bibitem [{\citenamefont {Spall}(1992)}]{spall_multivariate_1992}%
  \BibitemOpen
  \bibfield  {author} {\bibinfo {author} {\bibfnamefont {J.}~\bibnamefont
  {Spall}},\ }\href {\doibase 10.1109/9.119632} {\bibfield  {journal} {\bibinfo
   {journal} {IEEE Transactions on Automatic Control}\ }\textbf {\bibinfo
  {volume} {37}},\ \bibinfo {pages} {332–341} (\bibinfo {year}
  {1992})}\BibitemShut {NoStop}%
\bibitem [{\citenamefont {Lieb}\ \emph {et~al.}(1961)\citenamefont {Lieb},
  \citenamefont {Schultz},\ and\ \citenamefont {Mattis}}]{lieb_two_1961}%
  \BibitemOpen
  \bibfield  {author} {\bibinfo {author} {\bibfnamefont {E.}~\bibnamefont
  {Lieb}}, \bibinfo {author} {\bibfnamefont {T.}~\bibnamefont {Schultz}}, \
  and\ \bibinfo {author} {\bibfnamefont {D.}~\bibnamefont {Mattis}},\ }\href
  {\doibase 10.1016/0003-4916(61)90115-4} {\bibfield  {journal} {\bibinfo
  {journal} {Annals of Physics}\ }\textbf {\bibinfo {volume} {16}},\ \bibinfo
  {pages} {407} (\bibinfo {year} {1961})}\BibitemShut {NoStop}%
\bibitem [{\citenamefont {Pfeuty}(1970)}]{pfeuty_one-dimensional_1970}%
  \BibitemOpen
  \bibfield  {author} {\bibinfo {author} {\bibfnamefont {P.}~\bibnamefont
  {Pfeuty}},\ }\href {\doibase 10.1016/0003-4916(70)90270-8} {\bibfield
  {journal} {\bibinfo  {journal} {Annals of Physics}\ }\textbf {\bibinfo
  {volume} {57}},\ \bibinfo {pages} {79} (\bibinfo {year} {1970})}\BibitemShut
  {NoStop}%
\bibitem [{\citenamefont {Aleksandrowicz}\ \emph {et~al.}(2019)\citenamefont
  {Aleksandrowicz}, \citenamefont {Alexander}, \citenamefont {Barkoutsos},
  \citenamefont {Bello}, \citenamefont {Ben-Haim}, \citenamefont {Bucher},
  \citenamefont {Cabrera-Hernández}, \citenamefont {Carballo-Franquis},
  \citenamefont {Chen}, \citenamefont {Chen}, \citenamefont {Chow},
  \citenamefont {Córcoles-Gonzales}, \citenamefont {Cross}, \citenamefont
  {Cross}, \citenamefont {Cruz-Benito}, \citenamefont {Culver}, \citenamefont
  {González}, \citenamefont {Torre}, \citenamefont {Ding}, \citenamefont
  {Dumitrescu}, \citenamefont {Duran}, \citenamefont {Eendebak}, \citenamefont
  {Everitt}, \citenamefont {Sertage}, \citenamefont {Frisch}, \citenamefont
  {Fuhrer}, \citenamefont {Gambetta}, \citenamefont {Gago}, \citenamefont
  {Gomez-Mosquera}, \citenamefont {Greenberg}, \citenamefont {Hamamura},
  \citenamefont {Havlicek}, \citenamefont {Hellmers}, \citenamefont {Łukasz
  Herok}, \citenamefont {Horii}, \citenamefont {Hu}, \citenamefont {Imamichi},
  \citenamefont {Itoko}, \citenamefont {Javadi-Abhari}, \citenamefont
  {Kanazawa}, \citenamefont {Karazeev}, \citenamefont {Krsulich}, \citenamefont
  {Liu}, \citenamefont {Luh}, \citenamefont {Maeng}, \citenamefont {Marques},
  \citenamefont {Martín-Fernández}, \citenamefont {McClure}, \citenamefont
  {McKay}, \citenamefont {Meesala}, \citenamefont {Mezzacapo}, \citenamefont
  {Moll}, \citenamefont {Rodríguez}, \citenamefont {Nannicini}, \citenamefont
  {Nation}, \citenamefont {Ollitrault}, \citenamefont {O'Riordan},
  \citenamefont {Paik}, \citenamefont {Pérez}, \citenamefont {Phan},
  \citenamefont {Pistoia}, \citenamefont {Prutyanov}, \citenamefont {Reuter},
  \citenamefont {Rice}, \citenamefont {Davila}, \citenamefont {Rudy},
  \citenamefont {Ryu}, \citenamefont {Sathaye}, \citenamefont {Schnabel},
  \citenamefont {Schoute}, \citenamefont {Setia}, \citenamefont {Shi},
  \citenamefont {Silva}, \citenamefont {Siraichi}, \citenamefont {Sivarajah},
  \citenamefont {Smolin}, \citenamefont {Soeken}, \citenamefont {Takahashi},
  \citenamefont {Tavernelli}, \citenamefont {Taylor}, \citenamefont {Taylour},
  \citenamefont {Trabing}, \citenamefont {Treinish}, \citenamefont {Turner},
  \citenamefont {Vogt-Lee}, \citenamefont {Vuillot}, \citenamefont {Wildstrom},
  \citenamefont {Wilson}, \citenamefont {Winston}, \citenamefont {Wood},
  \citenamefont {Wood}, \citenamefont {Wörner}, \citenamefont {Akhalwaya},\
  and\ \citenamefont {Zoufal}}]{aleksandrowicz_qiskit:_2019}%
  \BibitemOpen
  \bibfield  {author} {\bibinfo {author} {\bibfnamefont {G.}~\bibnamefont
  {Aleksandrowicz}}, \bibinfo {author} {\bibfnamefont {T.}~\bibnamefont
  {Alexander}}, \bibinfo {author} {\bibfnamefont {P.}~\bibnamefont
  {Barkoutsos}}, \bibinfo {author} {\bibfnamefont {L.}~\bibnamefont {Bello}},
  \bibinfo {author} {\bibfnamefont {Y.}~\bibnamefont {Ben-Haim}}, \bibinfo
  {author} {\bibfnamefont {D.}~\bibnamefont {Bucher}}, \bibinfo {author}
  {\bibfnamefont {F.~J.}\ \bibnamefont {Cabrera-Hernández}}, \bibinfo {author}
  {\bibfnamefont {J.}~\bibnamefont {Carballo-Franquis}}, \bibinfo {author}
  {\bibfnamefont {A.}~\bibnamefont {Chen}}, \bibinfo {author} {\bibfnamefont
  {C.-F.}\ \bibnamefont {Chen}}, \bibinfo {author} {\bibfnamefont {J.~M.}\
  \bibnamefont {Chow}}, \bibinfo {author} {\bibfnamefont {A.~D.}\ \bibnamefont
  {Córcoles-Gonzales}}, \bibinfo {author} {\bibfnamefont {A.~J.}\ \bibnamefont
  {Cross}}, \bibinfo {author} {\bibfnamefont {A.}~\bibnamefont {Cross}},
  \bibinfo {author} {\bibfnamefont {J.}~\bibnamefont {Cruz-Benito}}, \bibinfo
  {author} {\bibfnamefont {C.}~\bibnamefont {Culver}}, \bibinfo {author}
  {\bibfnamefont {S.~D. L.~P.}\ \bibnamefont {González}}, \bibinfo {author}
  {\bibfnamefont {E.~D.~L.}\ \bibnamefont {Torre}}, \bibinfo {author}
  {\bibfnamefont {D.}~\bibnamefont {Ding}}, \bibinfo {author} {\bibfnamefont
  {E.}~\bibnamefont {Dumitrescu}}, \bibinfo {author} {\bibfnamefont
  {I.}~\bibnamefont {Duran}}, \bibinfo {author} {\bibfnamefont
  {P.}~\bibnamefont {Eendebak}}, \bibinfo {author} {\bibfnamefont
  {M.}~\bibnamefont {Everitt}}, \bibinfo {author} {\bibfnamefont {I.~F.}\
  \bibnamefont {Sertage}}, \bibinfo {author} {\bibfnamefont {A.}~\bibnamefont
  {Frisch}}, \bibinfo {author} {\bibfnamefont {A.}~\bibnamefont {Fuhrer}},
  \bibinfo {author} {\bibfnamefont {J.}~\bibnamefont {Gambetta}}, \bibinfo
  {author} {\bibfnamefont {B.~G.}\ \bibnamefont {Gago}}, \bibinfo {author}
  {\bibfnamefont {J.}~\bibnamefont {Gomez-Mosquera}}, \bibinfo {author}
  {\bibfnamefont {D.}~\bibnamefont {Greenberg}}, \bibinfo {author}
  {\bibfnamefont {I.}~\bibnamefont {Hamamura}}, \bibinfo {author}
  {\bibfnamefont {V.}~\bibnamefont {Havlicek}}, \bibinfo {author}
  {\bibfnamefont {J.}~\bibnamefont {Hellmers}}, \bibinfo {author} {\bibnamefont
  {Łukasz Herok}}, \bibinfo {author} {\bibfnamefont {H.}~\bibnamefont
  {Horii}}, \bibinfo {author} {\bibfnamefont {S.}~\bibnamefont {Hu}}, \bibinfo
  {author} {\bibfnamefont {T.}~\bibnamefont {Imamichi}}, \bibinfo {author}
  {\bibfnamefont {T.}~\bibnamefont {Itoko}}, \bibinfo {author} {\bibfnamefont
  {A.}~\bibnamefont {Javadi-Abhari}}, \bibinfo {author} {\bibfnamefont
  {N.}~\bibnamefont {Kanazawa}}, \bibinfo {author} {\bibfnamefont
  {A.}~\bibnamefont {Karazeev}}, \bibinfo {author} {\bibfnamefont
  {K.}~\bibnamefont {Krsulich}}, \bibinfo {author} {\bibfnamefont
  {P.}~\bibnamefont {Liu}}, \bibinfo {author} {\bibfnamefont {Y.}~\bibnamefont
  {Luh}}, \bibinfo {author} {\bibfnamefont {Y.}~\bibnamefont {Maeng}}, \bibinfo
  {author} {\bibfnamefont {M.}~\bibnamefont {Marques}}, \bibinfo {author}
  {\bibfnamefont {F.~J.}\ \bibnamefont {Martín-Fernández}}, \bibinfo {author}
  {\bibfnamefont {D.~T.}\ \bibnamefont {McClure}}, \bibinfo {author}
  {\bibfnamefont {D.}~\bibnamefont {McKay}}, \bibinfo {author} {\bibfnamefont
  {S.}~\bibnamefont {Meesala}}, \bibinfo {author} {\bibfnamefont
  {A.}~\bibnamefont {Mezzacapo}}, \bibinfo {author} {\bibfnamefont
  {N.}~\bibnamefont {Moll}}, \bibinfo {author} {\bibfnamefont {D.~M.}\
  \bibnamefont {Rodríguez}}, \bibinfo {author} {\bibfnamefont
  {G.}~\bibnamefont {Nannicini}}, \bibinfo {author} {\bibfnamefont
  {P.}~\bibnamefont {Nation}}, \bibinfo {author} {\bibfnamefont
  {P.}~\bibnamefont {Ollitrault}}, \bibinfo {author} {\bibfnamefont {L.~J.}\
  \bibnamefont {O'Riordan}}, \bibinfo {author} {\bibfnamefont {H.}~\bibnamefont
  {Paik}}, \bibinfo {author} {\bibfnamefont {J.}~\bibnamefont {Pérez}},
  \bibinfo {author} {\bibfnamefont {A.}~\bibnamefont {Phan}}, \bibinfo {author}
  {\bibfnamefont {M.}~\bibnamefont {Pistoia}}, \bibinfo {author} {\bibfnamefont
  {V.}~\bibnamefont {Prutyanov}}, \bibinfo {author} {\bibfnamefont
  {M.}~\bibnamefont {Reuter}}, \bibinfo {author} {\bibfnamefont
  {J.}~\bibnamefont {Rice}}, \bibinfo {author} {\bibfnamefont {A.~R.}\
  \bibnamefont {Davila}}, \bibinfo {author} {\bibfnamefont {R.~H.~P.}\
  \bibnamefont {Rudy}}, \bibinfo {author} {\bibfnamefont {M.}~\bibnamefont
  {Ryu}}, \bibinfo {author} {\bibfnamefont {N.}~\bibnamefont {Sathaye}},
  \bibinfo {author} {\bibfnamefont {C.}~\bibnamefont {Schnabel}}, \bibinfo
  {author} {\bibfnamefont {E.}~\bibnamefont {Schoute}}, \bibinfo {author}
  {\bibfnamefont {K.}~\bibnamefont {Setia}}, \bibinfo {author} {\bibfnamefont
  {Y.}~\bibnamefont {Shi}}, \bibinfo {author} {\bibfnamefont {A.}~\bibnamefont
  {Silva}}, \bibinfo {author} {\bibfnamefont {Y.}~\bibnamefont {Siraichi}},
  \bibinfo {author} {\bibfnamefont {S.}~\bibnamefont {Sivarajah}}, \bibinfo
  {author} {\bibfnamefont {J.~A.}\ \bibnamefont {Smolin}}, \bibinfo {author}
  {\bibfnamefont {M.}~\bibnamefont {Soeken}}, \bibinfo {author} {\bibfnamefont
  {H.}~\bibnamefont {Takahashi}}, \bibinfo {author} {\bibfnamefont
  {I.}~\bibnamefont {Tavernelli}}, \bibinfo {author} {\bibfnamefont
  {C.}~\bibnamefont {Taylor}}, \bibinfo {author} {\bibfnamefont
  {P.}~\bibnamefont {Taylour}}, \bibinfo {author} {\bibfnamefont
  {K.}~\bibnamefont {Trabing}}, \bibinfo {author} {\bibfnamefont
  {M.}~\bibnamefont {Treinish}}, \bibinfo {author} {\bibfnamefont
  {W.}~\bibnamefont {Turner}}, \bibinfo {author} {\bibfnamefont
  {D.}~\bibnamefont {Vogt-Lee}}, \bibinfo {author} {\bibfnamefont
  {C.}~\bibnamefont {Vuillot}}, \bibinfo {author} {\bibfnamefont {J.~A.}\
  \bibnamefont {Wildstrom}}, \bibinfo {author} {\bibfnamefont {J.}~\bibnamefont
  {Wilson}}, \bibinfo {author} {\bibfnamefont {E.}~\bibnamefont {Winston}},
  \bibinfo {author} {\bibfnamefont {C.}~\bibnamefont {Wood}}, \bibinfo {author}
  {\bibfnamefont {S.}~\bibnamefont {Wood}}, \bibinfo {author} {\bibfnamefont
  {S.}~\bibnamefont {Wörner}}, \bibinfo {author} {\bibfnamefont {I.~Y.}\
  \bibnamefont {Akhalwaya}}, \ and\ \bibinfo {author} {\bibfnamefont
  {C.}~\bibnamefont {Zoufal}},\ }\href {\doibase 10.5281/zenodo.2562110}
  {\enquote {\bibinfo {title} {Qiskit: {An} {Open}-source {Framework} for
  {Quantum} {Computing}},}\ } (\bibinfo {year} {2019})\BibitemShut {NoStop}%
\bibitem [{\citenamefont {Franchini}(2017)}]{franchini_introduction_2017}%
  \BibitemOpen
  \bibfield  {author} {\bibinfo {author} {\bibfnamefont {F.}~\bibnamefont
  {Franchini}},\ }\href {\doibase 10.1007/978-3-319-48487-7} {\emph {\bibinfo
  {title} {An {Introduction} to {Integrable} {Techniques} for
  {One}-{Dimensional} {Quantum} {Systems}}}},\ \bibinfo {series} {Lecture
  {Notes} in {Physics}}, Vol.\ \bibinfo {volume} {940}\ (\bibinfo  {publisher}
  {Springer International Publishing},\ \bibinfo {address} {Cham},\ \bibinfo
  {year} {2017})\BibitemShut {NoStop}%
\bibitem [{\citenamefont
  {Dillenschneider}(2008)}]{dillenschneider_quantum_2008}%
  \BibitemOpen
  \bibfield  {author} {\bibinfo {author} {\bibfnamefont {R.}~\bibnamefont
  {Dillenschneider}},\ }\href {\doibase 10.1103/PhysRevB.78.224413} {\bibfield
  {journal} {\bibinfo  {journal} {Physical Review B}\ }\textbf {\bibinfo
  {volume} {78}},\ \bibinfo {pages} {224413} (\bibinfo {year}
  {2008})}\BibitemShut {NoStop}%
\bibitem [{\citenamefont {Vidal}(2007)}]{vidal_entanglement_2007}%
  \BibitemOpen
  \bibfield  {author} {\bibinfo {author} {\bibfnamefont {G.}~\bibnamefont
  {Vidal}},\ }\href {\doibase 10.1103/PhysRevLett.99.220405} {\bibfield
  {journal} {\bibinfo  {journal} {Physical Review Letters}\ }\textbf {\bibinfo
  {volume} {99}} (\bibinfo {year} {2007}),\
  10.1103/PhysRevLett.99.220405}\BibitemShut {NoStop}%
\bibitem [{\citenamefont {Hastie}\ \emph {et~al.}(2017)\citenamefont {Hastie},
  \citenamefont {Friedman},\ and\ \citenamefont
  {Tisbshirani}}]{hastie_friedman_tisbshirani_2017}%
  \BibitemOpen
  \bibfield  {author} {\bibinfo {author} {\bibfnamefont {T.}~\bibnamefont
  {Hastie}}, \bibinfo {author} {\bibfnamefont {J.}~\bibnamefont {Friedman}}, \
  and\ \bibinfo {author} {\bibfnamefont {R.}~\bibnamefont {Tisbshirani}},\
  }\href@noop {} {\emph {\bibinfo {title} {The Elements of statistical
  learning: data mining, inference, and prediction}}}\ (\bibinfo  {publisher}
  {Springer},\ \bibinfo {year} {2017})\BibitemShut {NoStop}%
\bibitem [{\citenamefont {Hunter}(2007)}]{hunter_matplotlib:_2007}%
  \BibitemOpen
  \bibfield  {author} {\bibinfo {author} {\bibfnamefont {J.~D.}\ \bibnamefont
  {Hunter}},\ }\href {\doibase 10.1109/MCSE.2007.55} {\bibfield  {journal}
  {\bibinfo  {journal} {Computing in Science \& Engineering}\ }\textbf
  {\bibinfo {volume} {9}},\ \bibinfo {pages} {90} (\bibinfo {year}
  {2007})}\BibitemShut {NoStop}%
\bibitem [{\citenamefont {Garcia-Saez}\ and\ \citenamefont
  {Latorre}(2018)}]{garcia-saez_addressing_2018}%
  \BibitemOpen
  \bibfield  {author} {\bibinfo {author} {\bibfnamefont {A.}~\bibnamefont
  {Garcia-Saez}}\ and\ \bibinfo {author} {\bibfnamefont {J.~I.}\ \bibnamefont
  {Latorre}},\ }\href {http://arxiv.org/abs/1806.02287} {\bibfield  {journal}
  {\bibinfo  {journal} {arXiv:1806.02287 [cond-mat, physics:quant-ph]}\ }
  (\bibinfo {year} {2018})},\ \bibinfo {note} {arXiv: 1806.02287}\BibitemShut
  {NoStop}%
\end{thebibliography}

%

\onecolumngrid 
\newpage
\appendix
\section{Supplemental Material}
\subsection{Data augmentation for the XXZ model}

The XXZ Hamiltonian is symmetric with respect to spin flips and $xy$ plane rotations. Despite the fact that its ground state is non-degenerate, applying these symmetries to a VQE state produces a different state with the same energy, which is an equally valid data point. Thankfully, these actions can be easily performed on the checkerboard states without changing their structure.

Let us start considering rotation symmetry. This rotation is implementing by applying a $Z$ rotation to each qubit:
\begin{equation}
    \label{eq:rotate}
    U_{rot} = (e^{i \frac{\varphi}{2} Z})^{\otimes n}.
\end{equation}
In the ansatz we deloped, the two-qubit blocks precede $Z$ rotations. So, applying this symmetry amounts to changing the angles in the Z rotations of the last checkerboard layer by $\varphi$.

\begin{figure}
    \centering
    \mbox{
    \Qcircuit @C=1.0em @R=0.8em {
           \ket{0} \dots \qquad \quad & \gate{e^{-i \alpha \sigma_Z}} & \multigate{1}{e^{-i \gamma \sigma_{Z} \otimes \sigma_{Z}}} & \gate{e^{-i \delta \sigma_X}} & \gate{X} &\qw \\
           \ket{0} \dots \qquad \quad & \gate{e^{-i \beta \sigma_Z}} & \ghost{e^{-i \gamma \sigma_Z \otimes \sigma_Z}} & \gate{e^{-i \epsilon \sigma_X}} & \gate{X} & \qw \\
       }
    }
    \caption{Two-qubit entangler gate with additional X rotations applied.}
    \label{fig:entangler_and_x}
\end{figure}
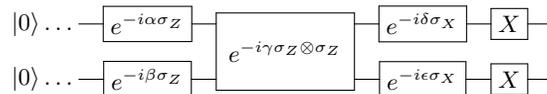

Somewhat more complicated is the application of spin flips. We consider spin flips as applying one of $X, Y,$ or $Z$ operations to all spin simultaneously. The Z spin flip is a special case of the $Z$ rotation(s). The $Y$ flip can be composed out of $X$ and $Z$ flips, so we only need consider the $X$ flip. Consider the quantum circuit in Fig. \ref{fig:entangler_and_x}.

Let us use the fact that $X$ and $Z$ Pauli matrices anticommute and push the $X$ gates to the front:
\begin{equation}
X e^{i \frac{\theta}{2}Z} = \cos \frac{\theta}{2} X + i \sin \frac{\theta}{2} X Z = (\cos \frac{\theta}{2} \eye - i \sin \frac{\theta}{2} Z) X = e^{- i \frac{\theta}{2}Z} X.
\end{equation}

Thus, if we push the $X$ gates to the left and invert the angles of the $Z$ rotations, the circuit remains invariant. Now, the next gate is the $Z \otimes Z$ rotation. $[X \otimes X, Z \otimes Z] = 0$, therefore the $X$ gates can go through the $R_{ZZ}$ gate without any changes. Finally, the $X$ gates merge with the $X$ rotations by incrementing the angle by $\pi / 2$. Thus, to augment the data with the spin-flipped states, one inverts the angles of the $Z$ rotations in the last layer, and increase the angles of the $X$ rotations in the last layer by $\pi / 2$.

\subsection{\add{Testing the processor on a model without structure}}

\add{As was pointed in the main text, simpler toy models may have simple classification criteria which do not require application of machine learning. In this section, we classify the solutions of a randomized model: 
$H(\alpha) = (1-\alpha) H_1 + \alpha H_2, \ \alpha \in [0, 1]$, where $H_1$ and $H_2$ are random Hermitian matrices pulled from a Gaussian unitary ensemble. We split the solutions in two classes: (i) $\alpha < 0.5$ and (ii) $\alpha > 0.5$. Then we run the optimization routine to train the learning circuit to discern between the two classes.}

\add{The approach was tested for 6 qubits, $n=100$, where $n_{train} = 70,\ n_{test}= 30$. The depth of the VQE circuit and the classifier circuit were both set to four layers.}

\begin{figure}
    \centering
    \includegraphics[width=0.8\textwidth]{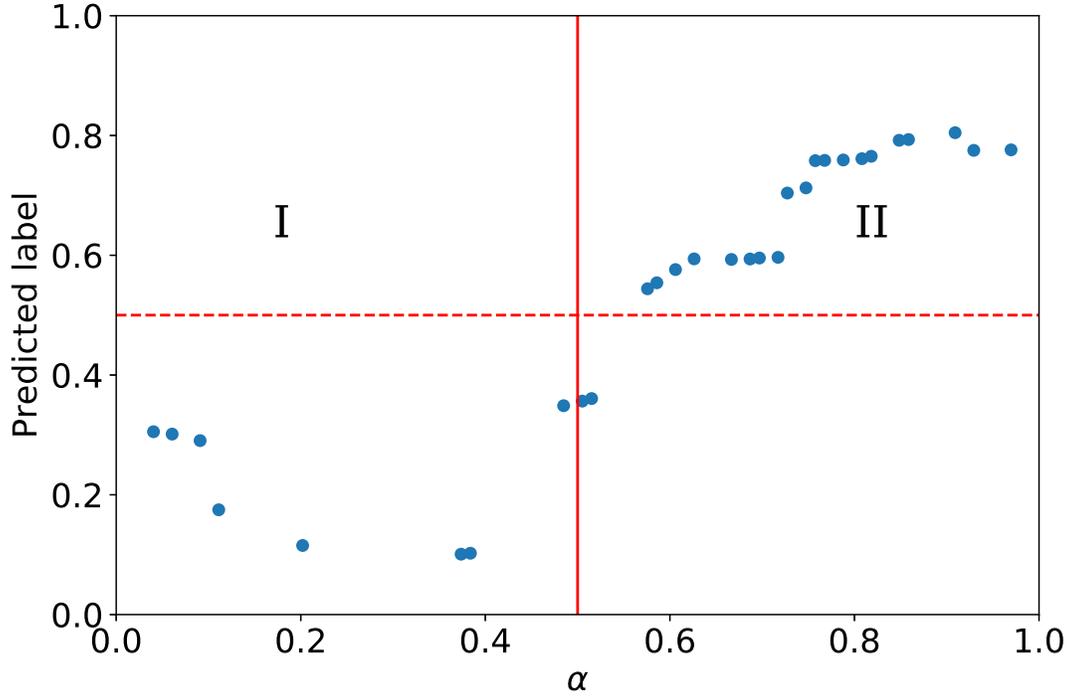}
    \caption{Results of the learning on the random Hamiltonians model.}
    \label{fig:learn_rand_ham}
\end{figure}

\add{The results are shown in Fig. \ref{fig:learn_rand_ham} . For this configuration, the accuracy of 93 \% was reached. This shows that the algorithm works even for such a low-structured problem, although the factors affecting the performance still require further investigation.}

{}

\subsection{\add{A detailed depiction of the quantum circuit}}

\begin{figure}
        \centering
        \includegraphics[width=0.8\textwidth]{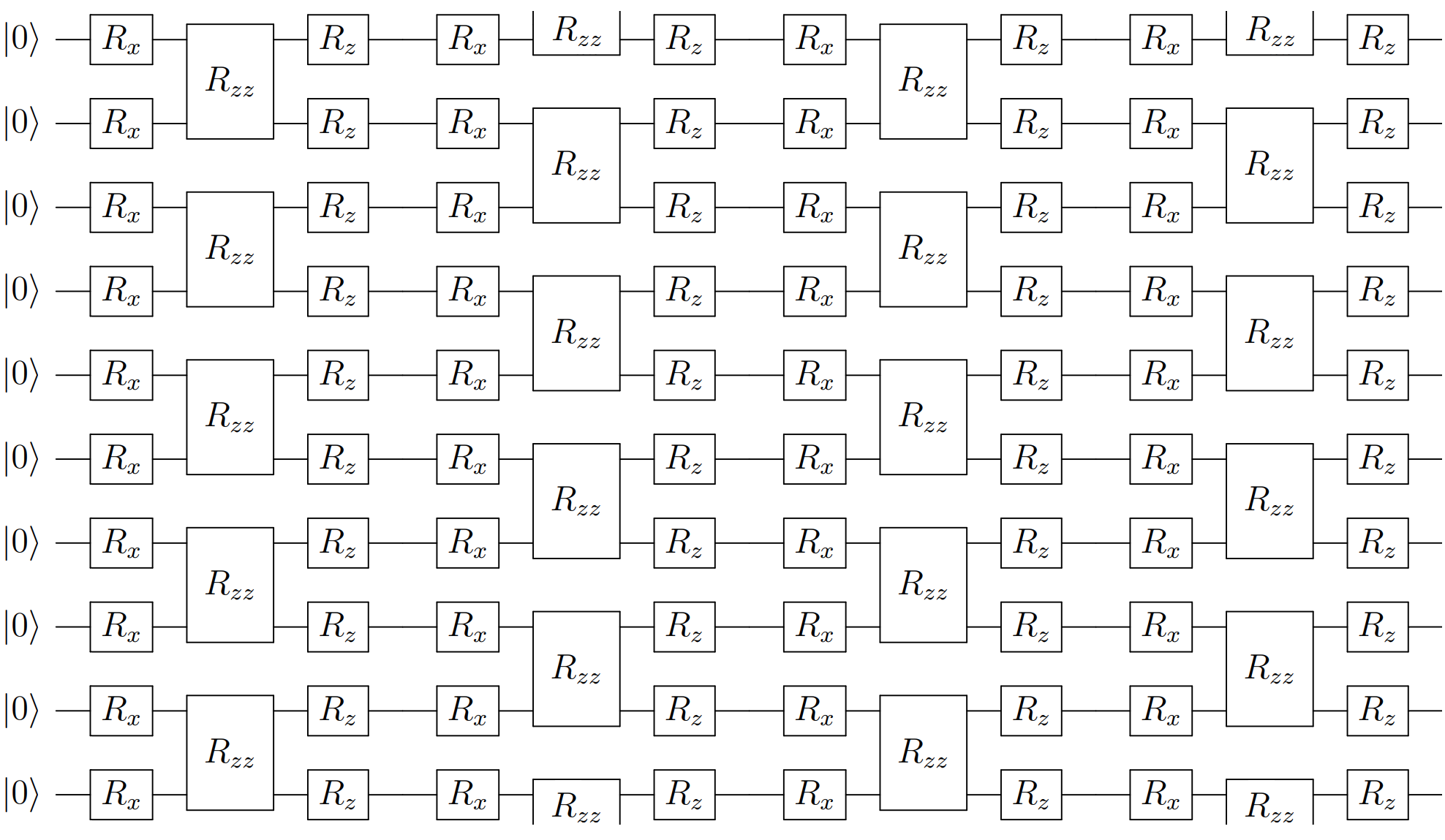}
        \caption{$75$-parameter ten-qubit circuit representing the classifier $U_{\mathrm{class}}$. Here $R_\sigma \equiv R_\sigma(\theta) = e^{-i \theta \sigma}$ and $R_{\sigma\sigma} \equiv R_{\sigma\sigma}(\theta) = e^{-i \theta \sigma \otimes \sigma}$, $\sigma \in \{\eye, X, Y, Z\}$.}
        \label{fig:ten-qubit_circ}
    \end{figure}

\add{Figure \ref{fig:ten-qubit_circ} shows the classifier circuit (4 layers) in full detail for 10 qubits. A full circuit (VQE + classifier) can be obtained by concatenating two copies of that circuit.}

    {}

\subsection{\add{On convergence of VQE with warm starts}}

\begin{figure}
    \centering
    \includegraphics[width=0.8\textwidth]{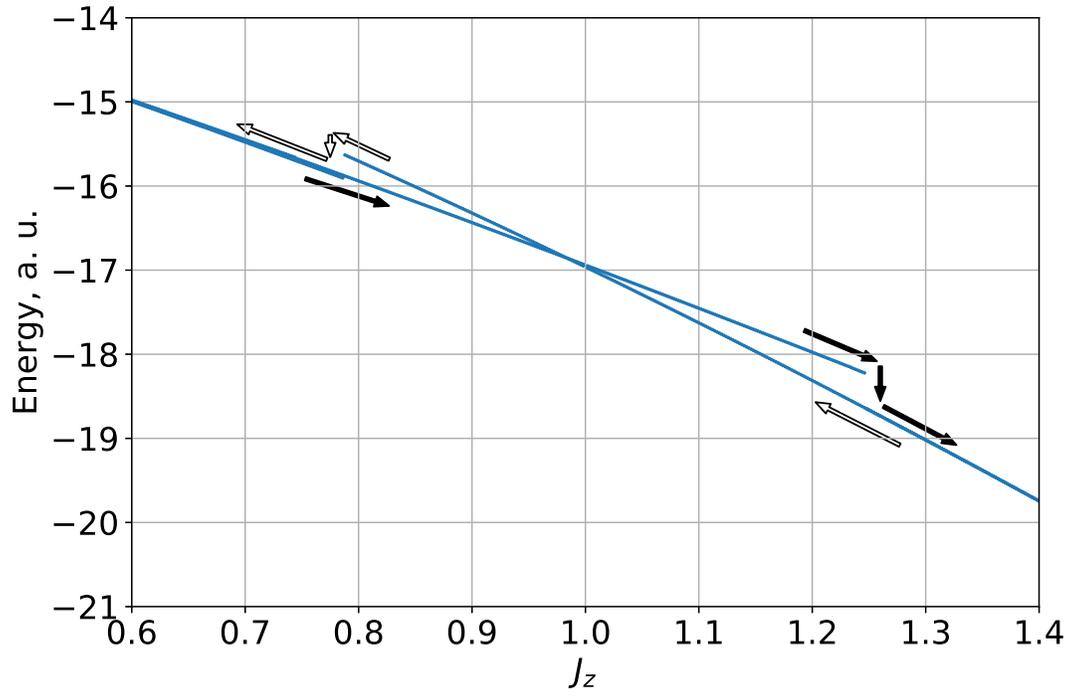}
    \caption{Ground state energy estimate for the XXZ model found in VQE sweeps. Filled (empty) arrows guide the eye along the ``up'' (``down'') sweep. Best solution out of two sweeps was subsequently used to train the classifier.}
    \label{fig:vqe_hyst}
\end{figure}

\add{To save time on VQE computations we used the previously found solution for $H(h)$ as a starting point for the VQE process on the next Hamiltonian $H(h + \Delta_h)$ (this approach is also known as adiabatic-assisted VQE [41]). Since the Hamiltonian is deformed only slightly, the previous point is a good guess for the new minimum.  Unfortunately, if during the deformation of the Hamiltonian this local minimum stops being a global one, the solver gets stuck in the wrong solution for a while. This is exactly what happens in the vicinity of the phase transition. Fig. \ref{fig:vqe_hyst} demonstrates the behavior of the VQE solution for the XXZ model. We ran VQE in two sweeps: $J_z$ swept from 0 to 2 in the ``up'' sweep and from 2 to 0 in the ``down'' sweep. As a result, VQE shows a hysteresis.}

{}


\end{document}